\documentclass[twocolumn,twoside,slac_two]{revtex4}
\usepackage{graphicx}
\usepackage{fancyhdr}

\usepackage{dcolumn}
\usepackage{amsmath}
\usepackage{epsfig}
\usepackage{multirow}
\usepackage{relsize}

\RequirePackage{xspace}
\usepackage{relsize}

\def\babar{\mbox{\slshape B\kern-0.1em{\smaller A}\kern-0.1em
    B\kern-0.1em{\smaller A\kern-0.2em R}}}
\def\Abar    {\kern 0.18em\overline{\kern -0.18em A}{}\xspace}
\def\Kbar    {\kern 0.18em\overline{\kern -0.18em K}{}\xspace}
\def\Bbar    {\kern 0.18em\overline{\kern -0.18em B}{}\xspace}
\def\Dbar    {\kern 0.18em\overline{\kern -0.18em D}{}\xspace}

\def\BB      {\ensuremath{B\Bbar}\xspace}
\def\Bz      {\ensuremath{B^0}\xspace}
\def\Bzb     {\ensuremath{\Bbar^0}\xspace}
\def\BzBzb   {\ensuremath{\Bz {\kern -0.16em \Bzb}}\xspace}
\def\Bu      {\ensuremath{B^+}\xspace}
\def\Bub     {\ensuremath{B^-}\xspace}

\def\BpBm    {\ensuremath{\Bu {\kern -0.16em \Bub}}\xspace}

\newcommand{\optbar}[1]{\shortstack{{\tiny (\rule[.4ex]{1em}{.1mm})}
  \\ [-.7ex] $#1$}}
\def\BorBbar    {\kern 0.18em\optbar{\kern -0.18em B}{}\xspace}
\def\DorDbar    {\kern 0.18em\optbar{\kern -0.18em D}{}\xspace}
\def\KorKbar    {\kern 0.18em\optbar{\kern -0.18em K}{}\xspace}

\def\pep2{PEP-II}
\mathchardef\Upsilon="7107
\def\Y#1S{\ensuremath{\Upsilon{(#1S)}}\xspace}

\def\FourS {\Y4S}

\def\btn {\ensuremath{B^{+} \to \tau^{+} \nu_\tau}\xspace}

\def\Vub {\ensuremath{V_{ub}}}

\def\btodx {\ensuremath{\Bub \to D^{(*)0} X^-}}

\def\eextra {\ensuremath{E_{\mathrm{extra}}}\xspace}

\def\tautoenunu {\ensuremath {\tau^+ \to e^+ \nu \nub}}

\def\tautomununu {\ensuremath {\tau^+ \to \mu^+ \nu \nub}}

\def\tautopinu {\ensuremath {\tau^+ \to \pi^+ \nub}}

\def\tautopipiznu {\ensuremath {\tau^+ \to \pi^+ \pi^{0} \nub}}
\def\tautopipiz {\ensuremath {\tau^+ \to \pi^+ \pi^{0} \nub}}

\def\tautothreepi {\ensuremath {\tau^+ \to \pi^+ \pi^{-} \pi^{+} \nub}}

\def\onlumi    {\ensuremath { 346  \invfb\ }}
\def\offlumi   {\ensuremath { 36.3 \invfb\  }}

\def\pipiz {\ensuremath { \pi^+ \pi^{0} }}

\def\invfb   {\ensuremath{\mbox{\,fb}^{-1}}\xspace}
\newcommand{\gev}{\ensuremath{\mathrm{\,Ge\kern -0.1em V}}\xspace}
\newcommand{\gevc}{\ensuremath{{\mathrm{\,Ge\kern -0.1em V\!/}c}}\xspace}
\newcommand{\gevcc}{\ensuremath{{\mathrm{\,Ge\kern -0.1em V\!/}c^2}}\xspace}
\newcommand{\mev}{\ensuremath{\mathrm{\,Me\kern -0.1em V}}\xspace}
\def\nub        {\ensuremath{\overline{\nu}}\xspace}
\def\mes        {\mbox{$m_{\rm ES}$}\xspace}
\def\KS    {\ensuremath{K^0_{\scriptscriptstyle S}}\xspace}
\def\piz   {\ensuremath{\pi^0}\xspace}
\def\Dstarz  {\ensuremath{D^{*0}}\xspace}

\def\Dz      {\ensuremath{D^0}\xspace}

\def\ra                 {\ensuremath{\rightarrow}\xspace}
\def\epem       {\ensuremath{e^+e^-}\xspace}
\def\uubar {\ensuremath{u\overline u}\xspace}
\def\ddbar {\ensuremath{d\overline d}\xspace}
\def\ssbar {\ensuremath{s\overline s}\xspace}
\def\BzBzb   {\ensuremath{\Bz {\kern -0.16em \Bzb}}\xspace}

\def\BpBm    {\ensuremath{\Bu {\kern -0.16em \Bub}}\xspace}

\def\cm   {\ensuremath{{\rm \,cm}}\xspace}
\def\B    {\ensuremath{B}\xspace}

\begin{document}

\begin{flushleft}
SLAC-PUB-12569\\
\babar-TALK-07/041
\\[10mm]
\end{flushleft}

\title{Hot Topics from the $\babar$ Experiment}
\author{A. V. Gritsan}
\affiliation{Johns Hopkins University, Baltimore, Maryland 21218, USA}

\begin{abstract}
With a sample of about 384 million $\BB$ pairs recorded with 
the $\babar$ detector, we search for the flavor-changing charged 
current transition $B^{\pm} \to \tau^{\pm} \nu$ and perform an 
amplitude analysis of the effective flavor-changing neutral current 
transition $B^\pm\to\varphi(1020)K^{*}(892)^\pm$.
We also extend our search for other $K^*$ final states
in the decay $B^0\to\varphi(1020)K^{*0}$
with a large $K^{*0}\to K^+\pi^-$ invariant mass.
Two samples of events with one reconstructed hadronic $B$ decay 
or one reconstructed semileptonic $B$ decay are selected, and
in the recoil a search for $B^{\pm} \to \tau^{\pm} \nu$ is performed.
We find a 2.6 $\sigma$ (3.2 $\sigma$ not including expected background 
uncertainty) excess in data which can be converted to a 
preliminary branching fraction central value of 
$\mathcal{B}(B^{\pm} \to \tau^{\pm} \nu)=
({1.20}^{+0.40+0.29}_{-0.38-0.30}\pm0.22)\times{10^{-4}}$.
With the decay $B^\pm\to\varphi(1020)K^{*}(892)^\pm$, 
twelve parameters are measured, where our measurements 
of ${f_L}=0.49\pm{0.05}\pm 0.03$, ${f_\perp}=0.21\pm 0.05\pm 0.02$,
and the strong phases point to the presence of a substantial
helicity-plus amplitude from a presently unknown source.
\end{abstract}

\maketitle

\thispagestyle{fancy}


\section{Introduction}

Until the new frontier energy is available at the LHC collider,
the Standard Model is being investigated in detail via
the rich flavor structure of the quarks and leptons.
The study of rare processes, such as flavor-changing neutral
and charged currents, already provides constraints on new physics
and will allow us to disentangle the flavor structure
of particles at the new energy scale once LHC energy is accessible.
Until 2008, flavor physics of the third-generation particles,
such as $b$-quark, offers best prospects for discoveries. 
For example, virtual ``loop'' transitions of the $B$
meson involve the heaviest presently known particles in the
FCNC loop and may reveal new particles through subtle effects.
In this paper, we report the most recent (``hot'') results 
on two rare processes $B^{\pm} \to \tau^{\pm} \nu$ and 
$B^\pm\to\varphi(1020)K^{*}(892)^\pm$ with the $\babar$ detector.

\begin{figure}[b]
  \begin{center}
    \includegraphics[width=0.99\linewidth]{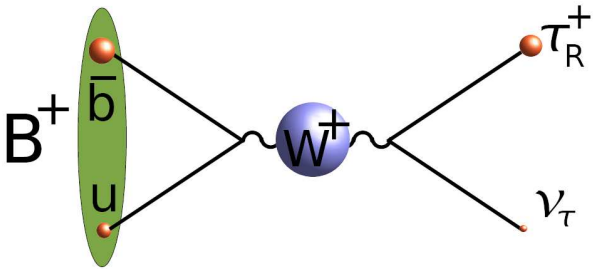}
    \caption{Feynman diagram describing the $B^{+} \to \tau^{+} \nu_\tau$ decay.}
    \label{fig:feyn-taunu}
  \end{center}
\end{figure}
\begin{figure}[b]
  \begin{center}
    \includegraphics[width=0.99\linewidth]{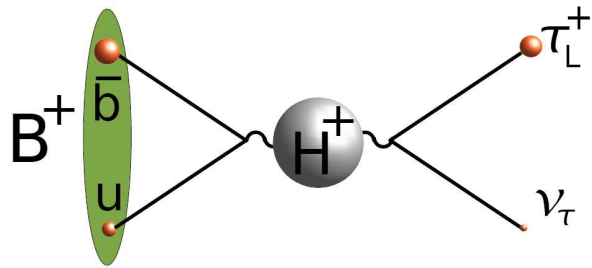}
\caption{Feynman diagram with potential charged Higgs contribution
to the $B^{+} \to \tau^{+} \nu_\tau$  decay.}
    \label{fig:feyn-taunu-higgs}
  \end{center}
\end{figure}

In the Standard Model (SM), the purely leptonic decay 
$B^{\pm} \to \tau^{\pm} \nu$
proceeds via quark annihilation into a $W^{+}$ boson,
as shown in Fig.~\ref{fig:feyn-taunu}.
The branching fraction is given by:
\begin{equation}
\label{eqn:br}
\mathcal{B}(B^{+} \rightarrow {\tau^+} \nu)=
\frac{G_{F}^{2} m^{}_{B}  m_{\tau}^{2}}{8\pi}
\left[1 - \frac{m_{\tau}^{2}}{m_{B}^{2}}\right]^{2}
\tau_{\Bu} f_{B}^{2} |\Vub|^{2},
\end{equation}
where
$\Vub$ is a quark mixing matrix element~\citep*{c,km},
$f_{B}$ is the $B$ meson decay constant,
$G_F$ is the Fermi constant,
$\tau_{\Bu}$ is the $\Bu$ lifetime, and
$m^{}_{B}$ and $m_{\tau}$ are the $\Bu$ meson and $\tau$ masses.
Physics beyond the SM, such as two-Higgs doublet models,
could enhance or suppress $\mathcal{B}(\btn)$ through the
introduction of a charged Higgs boson~\cite{higgs,Isidori2006pk,Akeroyd2007eh},
see Fig.~\ref{fig:feyn-taunu-higgs}.
Branching fraction in Eq.~(\ref{eqn:br}) is modified by a factor:
\begin{equation}
\label{eqn:brmod}
\left[
1 - { {\tan^2\beta}~\frac{m_{B^+}^{2}}{m_{H^+}^{2}}  }
\right]^{2}~,
\end{equation}
where $\tan\beta$ is the ratio of vacuum expectation values of the two
Higgs doublets.
Using theoretical calculations
of $f_B$ from lattice QCD and experimental measurements of $|\Vub|$
from semileptonic $B$ decays, this purely leptonic $B$-decay can be used
to constrain the parameters of theories beyond the SM.
Or, assuming that SM processes dominate and using the value
of $|\Vub|$ determined from semileptonic $B$-decays,
purely leptonic decays provide a clean experimental method of
measuring $f_B$ precisely.

The branching fractions for $B^+\to\mu^+\nu$ and $B^+\to e^+\nu$ are suppressed
by factors of $\sim 5\times 10^{-3}$ and $\sim 10^{-7}$ with respect to $\btn$. 
The SM estimate of the branching fraction for $\btn$,
using $|\Vub| = (4.31 \pm 0.30)\times 10^{-3}$~\cite{pdg2006} and
$f_{B} = 0.216 \pm 0.022$ GeV~\citep*{fb} in Eq.~\ref{eqn:br}
is $(1.6 \pm 0.4)\times 10^{-4}$.
However, a search for $\btn$ is experimentally challenging
due to the large missing momentum from multiple neutrinos, which makes
the signature less distinctive than in the other leptonic modes.
In a previously published analysis using a sample of
$223 \times 10^6$ $\FourS$ decays,
the \babar\ collaboration set an upper limit of
$\mathcal{B}(\btn) < 1.8 \times 10^{-4} \,
\textrm{ at the 90\% confidence level (CL)}$~\citep{babar-prd-btn}.
The Belle Collaboration has reported evidence from a search for this 
decay where the branching fraction was measured to be
$\mathcal{B}(\btn) = 
(1.79^{+0.56}_{-0.49}(\mbox{stat.})^{+0.46}_{-0.51}(\mbox{syst})) 
\times 10^{-4}$~\citep{belle}.


\begin{figure}[t]
  \begin{center}
    \includegraphics[width=0.99\linewidth]{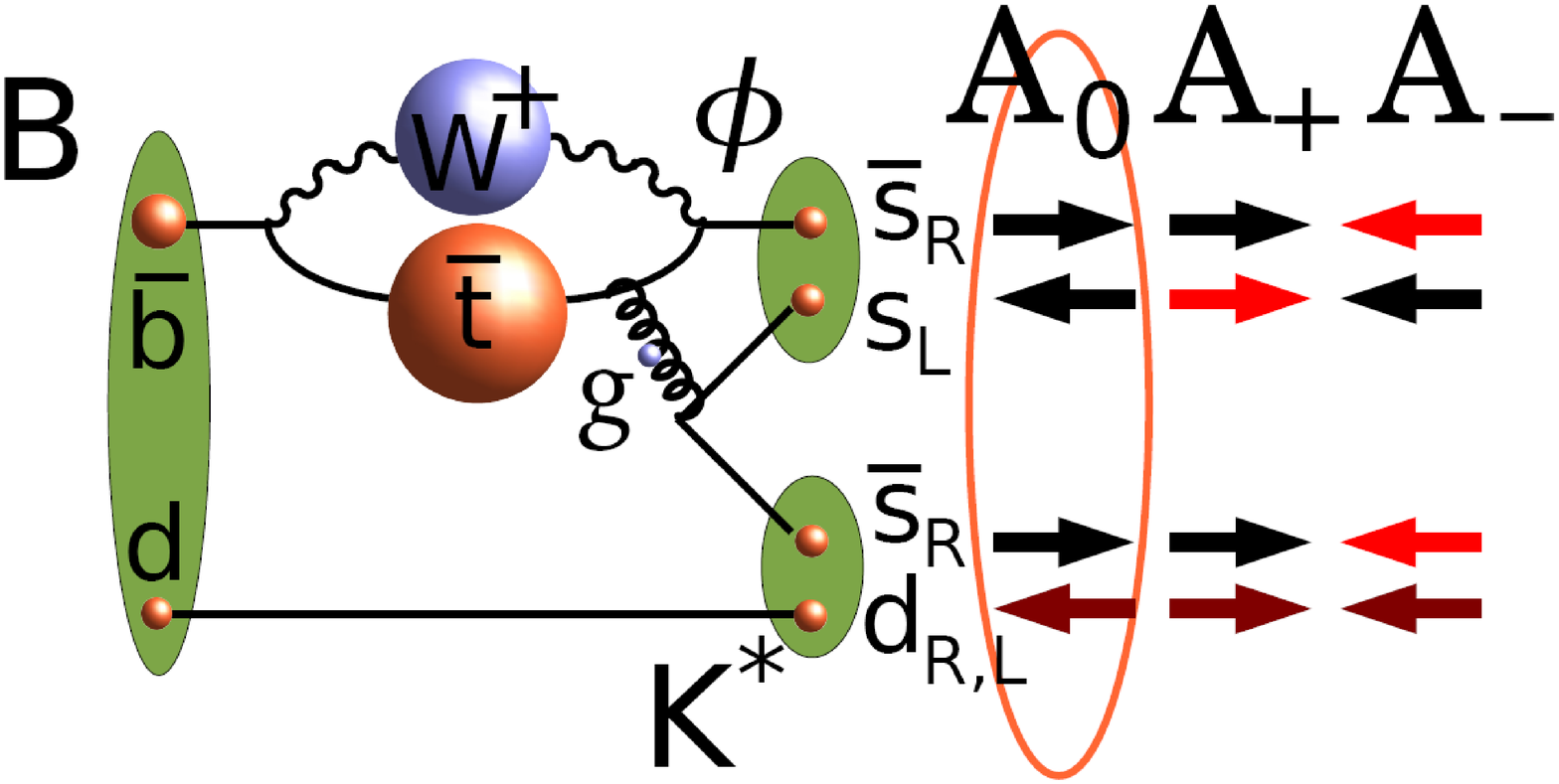}
\caption{Feynman diagram describing the $B\to\varphi K^{*}$  decay.
Due to the $(V-A)$ nature of the weak interactions, helicity conservation
in the strong interactions, and quark-spin suppression shown in the
diagram, we expect the hierarchy $A_{0}\gg A_{+1}\gg A_{-1}$.
}
    \label{fig:feyn-phikst}
  \end{center}
\centerline{
\setlength{\epsfxsize}{0.99\linewidth}\leavevmode\epsfbox{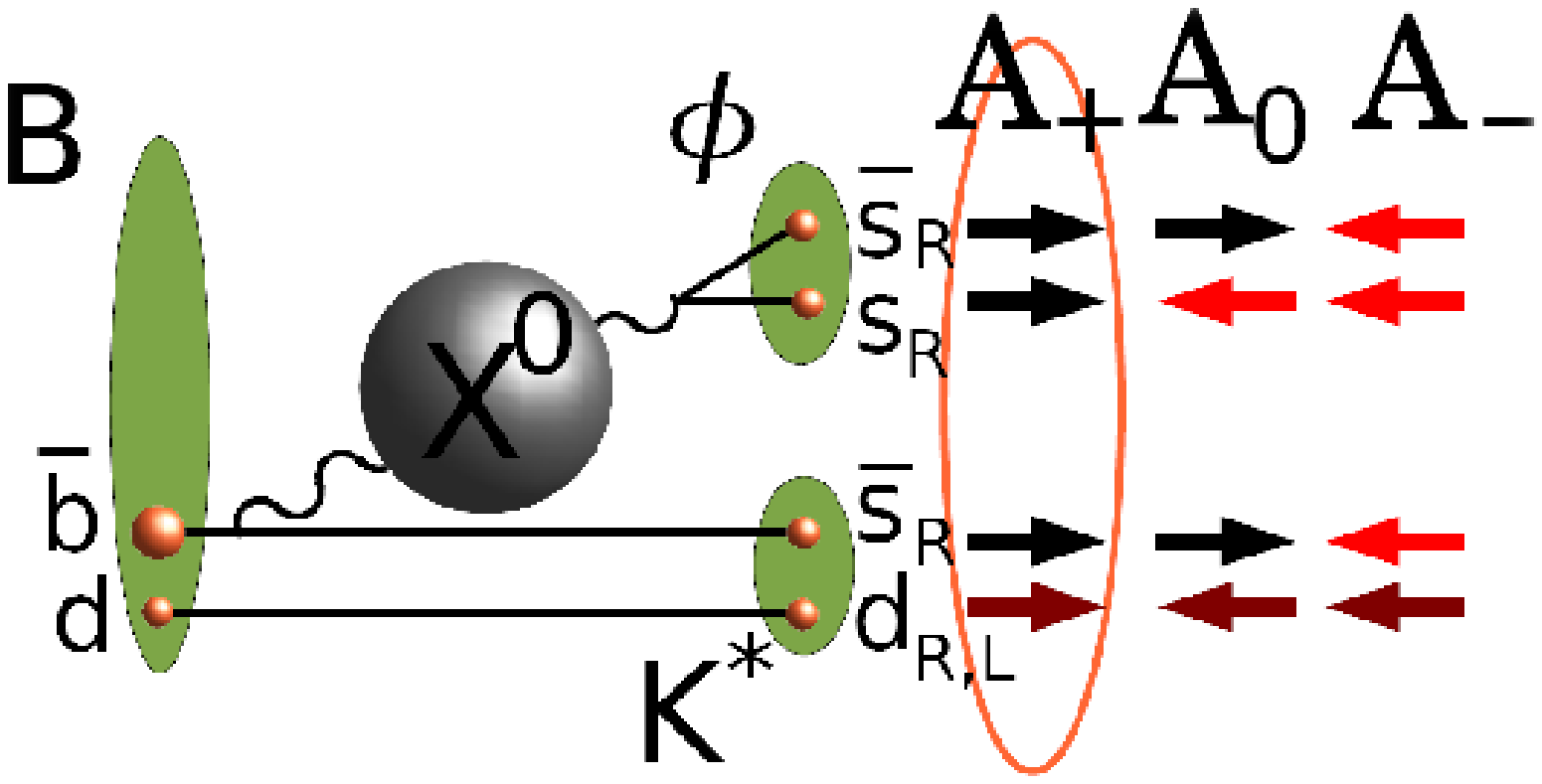}
}
\centerline{
\setlength{\epsfxsize}{0.99\linewidth}\leavevmode\epsfbox{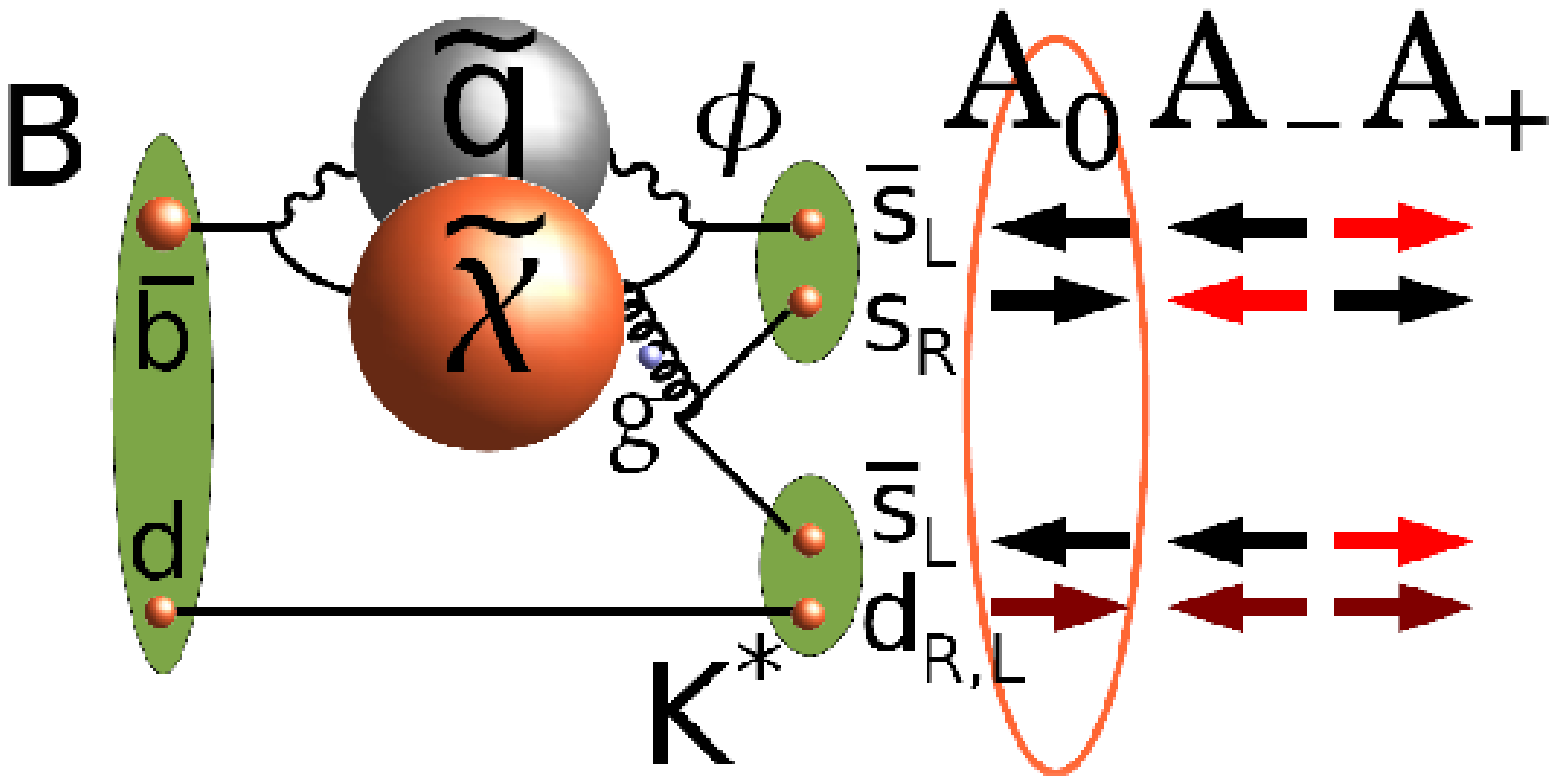}
}
\caption{Feynman diagram describing the $B\to\varphi K^{*}$  decay.
Scalar interaction may modify the hierarchy to $A_{+1}\gg A_{0}\gg A_{-1}$,
while supersymmetric interaction with $(V+A)$ couplings could produce
$A_{0}\gg A_{-1}\gg A_{+1}$.
}
\label{fig:feyn-phikst-new}
\end{figure}

A vector-vector $B$-meson decay, such as $B\to \varphi K^{*}$,
is characterized by three complex helicity amplitudes $A_{1\lambda}$
which correspond to helicity states $\lambda=-1,0,+1$ of the
vector mesons. The $A_{10}$ amplitude is expected to dominate~\cite{bvv1}
due to the $(V-A)$ nature of the weak interactions and helicity conservation
in the strong interactions, see Fig.~\ref{fig:feyn-phikst}.
A large fraction of transverse polarization observed by $\babar$
and confirmed by Belle, along with more recent
measurements of polarization in rare vector-vector
$B$ meson decays $B\to\phi K^*$ and $\rho K^*$,
indicate a significant departure from the
expected predominance of the longitudinal amplitude~\cite{babar:vv, belle:phikst,
belle:rhokst, babar:rhokst, babar:vt}.
The rate, polarization, and $C\!P$ measurements of $B$ meson decays
to particles with nonzero spin are sensitive to both strong and
weak interaction dynamics and are discussed
in a recent review~\cite{bvvreview2006,pdg2006}.

The polarization anomaly in vector-vector $B$ meson decays
suggests other contributions to the decay amplitude, previously
neglected. This has motivated a number of proposed
contributions from physics beyond the standard model~\cite{nptheory}.
Depending on New Physics model, hierarchy of decay amplitudes
could be modified, as shown in Fig.~\ref{fig:feyn-phikst-new}.
In addition, there are new mechanisms within the
standard model which have been proposed to address the anomaly,
such as annihilation penguin~\cite{smtheory} or electroweak penguin, 
or QCD rescattering~\cite{qcdtheory},

In order to distinguish the models,
the $\babar$ experiment extended the study
of the $B^0\to\phi K^{*0}$ decays with the tensor ($J^P = 2^+$),
vector ($J^P = 1^-$), and scalar ($J^P = 0^+$) $K^{*0}$~\cite{babar:vt}.
The vector-tensor results are in agreement with
quark spin-flip suppression~\cite{bvv1} and $A_0$ amplitude dominance,
whereas the vector-vector mode contains
substantial $A_{+1}$ amplitude, corresponding to anomalously
large transverse polarization.

We now investigate the polarization puzzle with a full
amplitude analysis of the $B^\pm\to\varphi K^{*}(892)^\pm$ decay.
In this paper, we report twelve independent parameters for the
three $B^+$ and three $B^-$ decay amplitudes, six of which are
presented for the first time.
Moreover, we use the dependence on the $K\pi$ invariant mass
of the interference between the $J^P=1^-$ and $0^+$
$(K\pi)^{\pm}$ components~\cite{babar:vt, Aston:1987ir, jpsikpi}
to resolve the discrete ambiguity between the
$A_{1 +1}$ and $A_{1 -1}$ helicity amplitudes.


\section{The $\babar$ Detector}

We use a sample of $383.6\pm 4.2$ million $\FourS\to\BB$ events
collected with the \babar\ detector~\cite{babar} at the PEP-II $e^+e^-$
asymmetric-energy storage rings. The $e^+e^-$ center-of-mass energy $\sqrt{s}$
is equal to $10.58$ GeV.
The sample corresponds to an integrated
luminosity of \onlumi at the \FourS\ resonance (on-resonance)
and \offlumi taken at $40\mev$ below the $B\bar{B}$ production threshold
(off-resonance) which is used to study background from
$e^+e^-\to f\bar{f}$ ($f = u, d, s, c, \tau$) continuum events.
The detector components used in this analysis are the tracking system
composed of a five-layer silicon vertex detector and a 40-layer drift chamber (DCH),
the Cherenkov detector for charged $\pi$--$K$ discrimination, a CsI calorimeter
(EMC) for photon and electron identification, and an
18--layer flux return (IFR) located outside of the 1.5~T solenoidal coil
and instrumented with resistive plate chambers for muon
and neutral hadron identification.
For the most recent 133 \invfb of data, a portion of the
resistive plate chambers
has been replaced with limited streamer tubes.

A GEANT4-based \cite{geant} Monte Carlo (MC)
simulation is used to model signal efficiencies and physics backgrounds.
The $\tau$ lepton decay is modeled using EvtGen \cite{evtgen}.
Beam-related background and detector
noise from data are overlaid on the simulated events.
Simulation samples equivalent to approximately three times the accumulated 
data are used to model \BB\ events, and samples equivalent to approximately
1.5 times the accumulated data are used to model continuum events.
We determine selection efficiencies for signal events using a
MC simulation where one $B^+$ meson decays to $\tau^+\nu$,
while the other is allowed to decay into any final state.


\section{\boldmath $B^{\pm} \to \tau^{\pm} \nu$}

\begin{figure}[t]
  \begin{center}
    \includegraphics[width=0.99\linewidth]{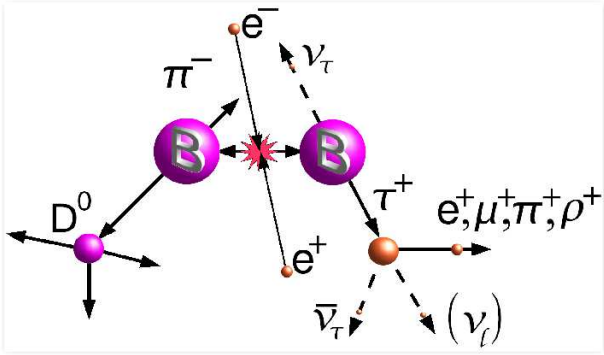}
    \caption{Diagram of he $B^{\pm} \to \tau^{\pm} \nu$ reconstruction technique.
}
    \label{fig:taunu-cartoon}
  \end{center}
\end{figure}
First we discuss the new preliminary analysis of the
$B^{\pm} \to \tau^{\pm} \nu$ decay.
Due to the presence of multiple neutrinos, the \btn decay mode
lacks the kinematic constraints that are usually exploited in $B$ decay
searches in order to reject both continuum and $\BB$ backgrounds.
The strategy adopted for this analysis is to reconstruct exclusively
the decay of one of the $B$ mesons in the event, referred to as the
``tag'' $B$. The remaining particle(s) in the event 
(the ``recoiling system'') are assumed to
come from the other $B$ and are compared with the signature
expected for \btn, see Fig.~\ref{fig:taunu-cartoon}.
There are two ``tag'' methods used in this analysis:
full reconstruction of a hadronic $B$ meson decay final state
or partial reconstruction of a semileptonic decay. We will discuss
in detail the new results with the former method and then briefly
show updated results of the latter method with slightly improved 
precision compared to last year. Then we will combine the two
results.

\subsection{Analysis Method}


In order to avoid experimenter bias, the
signal region in data is  blinded until the final yield
extraction is performed.
The \btn signal is searched for in
both leptonic and hadronic $\tau$ decay modes constituting approximately
71\% of the total $\tau$ decay width:
$\tautoenunu$, $\tautomununu$, $\tautopinu$, and $\tautopipiznu$.
We do not consider the $\tautothreepi$ mode since we found
it to be dominated by background events.


The tag $B$ candidate is reconstructed in the set of hadronic $B$ decay modes
\btodx~\cite{cc}, where $X^-$ denotes a system of
charged and neutral hadrons with total charge $-1$,
composed of $n_1\pi^{\pm}\, n_2K^{\pm}\, n_3\KS\,  n_4\piz$, where $n_1 + n_2 \leq
5$,  $n_3  \leq  2$,  and  $n_4  \leq  2$.

The selected sample of tag $B$ candidates is used as normalization for the
determination of the branching fraction.
We  reconstruct $\Dstarz \ra \Dz\piz, \Dz\gamma$;
$\Dz\ra K^-\pi^+$,$K^-\pi^+\piz$, $K^-\pi^+\pi^-\pi^+$,  
$\KS\pi^+\pi^-$; and  $\KS \ra \pi^+\pi^-$.
The kinematic consistency of tag $B$ candidates
is checked with two variables,
the beam energy-substituted mass $\mes = \sqrt{s/4 -
\vec{p}^{\,2}_B}$ and the energy difference
$\Delta E = E_B - \sqrt{s}/2$. Here $\sqrt{s}$ is the total
energy in the \FourS center-of-mass (CM) frame, and $\vec{p}_B$ and $E_B$
denote, respectively, the momentum and energy of the tag $B$ candidate in the CM
frame.  The resolution on $\Delta E$ is measured to be 
$\sigma_{\Delta E}=10-35\mev$, depending on
the decay mode, and we require $|\Delta E|<3\sigma_{\Delta E}$.
For each reconstructed $B$ decay mode, its purity
${\cal P}$ was estimated
as the ratio of the number of peaking events with 
\mes$>5.27$\gevc to the total number of events in the same range,
and is evaluated on on-resonance data.
In the event of multiple tag $B$ candidates being reconstructed, 
the one with the best purity ${\cal P}$ is selected.

\begin{figure}[t]
  \begin{center}
    \includegraphics[width=0.99\linewidth]{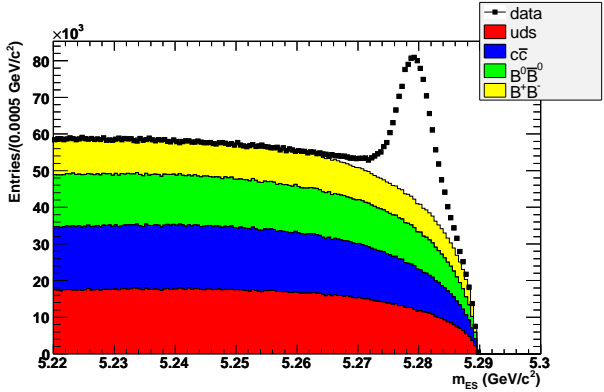}
    \caption{Cumulative distribution of the energy substituted mass \mes
      of the tag B candidates in data.
      The non peaking background components are added one on top of
      the other. From top to bottom are $B^+B^-$,
      $B^0\bar{B}^0$, uds and $c\bar{c}$.}
    \label{fig:mesfit}
  \end{center}
\end{figure}

The background consists of  $\epem\ra q\bar q \ (q=u,d,s,c)$ events and of
other $\FourS\to\BzBzb$ or \BpBm decays,
in which the tag $B$ candidate is mistakenly reconstructed from particles
coming from both $B$ mesons in the event.
To significantly reduce the $\epem\ra q\bar q$ background we require
the angle $\theta_{TB}^*$, defined in the CM frame, between the thrust
axis~\cite{thrust} of the tag $B$ candidate and the thrust axis of all the
charged and neutral reconstructed candidates
in the event excluding the ones that form the \Bu , to satisfy the requirement
$|\cos{\theta_{TB}^*}|<0.9$.

In order to determine the number of correctly reconstructed 
$\Bu$ we classify the background events in four
categories: $\epem\ra c\bar c$;
$\epem\ra \uubar,\;\ddbar,\;\ssbar$; \BzBzb; and \BpBm.
The shapes of these background distributions are taken from MC simulation.
The normalizations of the  $\epem\ra c\bar c$ and $\epem\ra \uubar,\;\ddbar,\;\ssbar$ 
backgrounds are taken from
off-resonance data, scaled by the luminosity.
The normalization of the \BzBzb, \BpBm components are instead
obtained by means of  a $\chi^2$ fit to the \mes distribution in
a sideband region ($5.22\gevcc<\mes<5.26\gevcc$).
The background contamination in the signal region ($\mes>5.27\gevcc$) 
is extrapolated from the fit and
subtracted from the data. We estimate the total number of tagged $B$'s in the data as
$N_{B} = (5.92 \pm 0.11_{\rm stat.}) \times 10^{5}$.
Fig.~\ref{fig:mesfit} shows the tag $B$ candidate \mes\ distribution in data compared
with MC, and the obtained background subtracted \mes\ distributions in data.


After the reconstruction of the tag $B$ meson, a selection 
is applied on the recoiling system in order to
enhance the sensitivity to \btn decays.
We require the presence of a well reconstructed charged track (signal track)
with electric charge opposite to the tag $B$. The signal track is required to have at least
12 hits in the DCH, its momentum transverse to the
beam axis, $p_{\rm{T}}$, is required to be greater than 0.1$\gevc$, and
its point of closest approach to the interaction point must be
less than 10.0\cm along the beam axis and less than 1.5\cm transverse
to it.

The $\tau$ lepton is identified in the four decay modes: $\tautoenunu$, 
$\tautomununu$, $\tautopinu$, and \tautopipiznu .
Particle identification criteria on the signal track are used to separate the four
categories.
The  \tautopipiznu\ sample is obtained by associating the signal track, identified
as pion, to a $\pi^0$ reconstructed from a pair  of neutral clusters with invariant mass
between 0.115 and 0.155 GeV/$c^2$ and total energy greater than 250 MeV. 
In case of multiple  \pipiz\ candidates the one with largest momentum 
$p^*_{\pi^+\piz}$ is chosen.

Most of the background due to continuum events and incorrectly reconstructed 
tag $B$ candidates (combinatorial) is rejected
by requiring a mode dependent cut on $|cos\theta^*_{TB}|$.
Most of the remaining sources of background  consists of $\BpBm$ events 
in which the tag $B$ meson was correctly reconstructed and the
recoil contains one track and additional particles which are not 
reconstructed by the tracking detectors and calorimeter.
From MC simulation we observe that most of this background 
is from semileptonic $B$ decays.

We define the discriminating variable \eextra as the sum of the energies 
of the neutral clusters not associated
with the tag $B$ or with the signal $\pi^0$ of the \tautopipiz\ mode.
For neutral clusters contributing to \eextra we need to
determine mode-dependent minimum energy thresholds, shown in 
Table~\ref{tab:minimumcl}, in order to ensure
a good data-MC agreement at the preselection level, which is a looser
selection using only particle identification criteria and a moderate
cut on the candidate's momentum.
Signal events tend to peak at low \eextra values whereas background 
events, which contain additional sources
of neutral clusters, are distributed toward higher \eextra values.
\begin{table}[!tbhp]
  \caption{EMC barrel and EMC endcap minimum cluster energies (\mev)
for electrons, muons and pions samples}
  \begin{center}
    \begin{tabular}{|l|c|c|} 
      \hline
      Sample    &    barrel &   endcap  \\ 
      \hline
      electrons &    65     &   70  \\
      muons     &    50     &   55  \\
      pions     &    50     &   70  \\
      \hline
    \end{tabular}
  \end{center}
  \label{tab:minimumcl}
\end{table}

Other variables used to discriminate between signal and  background are the CM
momentum of the signal candidates, the multiplicity of
extra charged track(s) and $\pi^0$(s) in the recoil and
the direction of the missing momentum four-vector in the CM frame.
For the  $\tautopipiznu$ mode we exploit the presence of the 
$\pi^0$ in the final state and the dominance of the decay through 
the $\rho$ resonance by means of the combined quantity  $s_{\rho}$:
\begin{equation}
\label{eq:rhoQF}
s^2_{\rho} = (\frac{m_{\pi\pi^0}-m^{PDG}_\rho}{\Gamma^{PDG}_\rho})^2
           + (\frac{m_{\gamma\gamma}-m^{PDG}_{\pi^0}}{\sigma_{\pi^0}})^2
\end{equation}
where $m_{\pi\pi^0}$ is the reconstructed invariant mass of the \pipiz\ 
candidate, $m_{\gamma\gamma}$
is the the reconstructed invariant mass of the $\pi^0$ candidate, 
$m^{PDG}_\rho$ and $\Gamma^{PDG}_\rho$ are the
nominal values~\cite{pdg2006} for the $\rho$ mass and width, 
$m^{PDG}_{\pi^0}$ is the nominal $\pi^0$ mass and $\sigma_{\pi^0}$ $=$ 8 MeV
is the experimental resolution on the $\pi^0$ mass determined from data.

We optimize the selection on the \BpBm
MC  and signal MC for the best $s/\sqrt{s+b}$, where $s$ is the expected signal and
$b$ is the expected background from \BpBm events,
in the hypothesis of a branching fraction
of $1 \times 10^{-4}$, for each mode separately. 
The optimization procedure is performed simultaneously on
all the discriminating variables in order to take into account any correlations.
For each discriminating variable $\alpha_i$ we choose 
a discrete number $N_{\alpha_i}$ of possible selection criteria.
We build N= $N_{\alpha_1}\times N_{\alpha_2} \times .... N_{\alpha_k}$ possible selections
by combining all the possible cut values for
different variables. We include \eextra in the optimization,
which defines the optimal signal window.
We choose the selection corresponding to the best value of $s/\sqrt{s+b}$.
The optimized selection criteria are reported in Table \ref{tab:selcuts}.

\begin{table}[!tbhp]
\caption{Selection criteria optimized for each $\tau$ decay mode. }
\begin{center}
\begin{tabular}{|l|rrrr|} 
      \hline
      Variable  & $e^+\nu\nu$ & $\mu^+\nu\nu$ & $\pi^+\bar\nu$ & $\pi^+\pi^0\bar\nu$ \\
      \hline
      \eextra (\gev)             &  $< 0.160$ & $< 0.100$ & $< 0.230$   & $< 0.290$ \\
      \piz multiplicity          &     0         &        0     &    $\leq 2$        &  n.a.       \\
      Track multipl. & 1             &  1           &    $\leq 2$        &  1          \\
      $|cos\theta^*_{TB}|$  & $\leq 0.9$     & $\leq 0.9$ & $\leq 0.7$ & $\leq 0.7$ \\
      $p^{*}_{\rm{trk}} (\gevc)$ & $<1.25$  &  $<1.85$  & $>1.5$   &  n.a.       \\
      $cos \theta^*_{\rm{miss}}$& $<0.9$           &  n.a         & $<0.5$             & $<0.55$     \\
      $p^{*}_{\pipiz} (\gevc) $       & n.a.          &  n.a.        & n.a.            & $>1.5$ \\
      $ \rho $ quality           & n.a.          &  n.a.        & n.a.           & $<2.0$         \\
      $ E_{\piz} $~(\gev)        & n.a.          &  n.a.        & n.a.        & $>0.250$         \\
      \hline
\end{tabular}
\end{center}
\label{tab:selcuts}
\end{table}

We compute the efficiency as the ratio of the number of signal MC events
passing the selection criteria and the number of events
that have a \mes peaking tag \B\ candidate, in the signal region $\mes > 5.27 \gev$.
We evaluate the efficiencies on a signal MC sample which is distinct
from the sample used in the optimization procedure.
A small cross-feed is present in some modes and it is taken
into account in the computation of the total efficiency.

The total efficiency for each selection is:
\begin{equation}
  \varepsilon_i = \sum_{j=1}^{n_{dec}} \varepsilon_i^j f_j \,\, ,
\end{equation}
where $\varepsilon_i^j$ is the efficiency of the selection $i$
for the $\tau$ decay mode $j$, $n_{\rm{dec}}=7$ is the number of of $\tau$
decay modes and $f_j $ are the
fractions of the $\tau$ decay mode as estimated from the signal MC 
sample with a reconstructed B. Table~\ref{tab:eff} shows 
the estimated efficiencies.

\begin{table}[!tbhp]
  \caption{ 
Selection efficiency ($\%$) for the $\tau$ decay modes.
The last row shows the total efficiency including
small cross-feed from other $\tau$ decays (not shown explicitly)
and weighted by the decay abundance at the tag
selection level. The errors are statistical only. Note that
mode dependent \eextra selection is applied as reported 
in Table~\ref{tab:selcuts}.}
    \begin{center}
    \begin{tabular}{|l|c|} 
\hline 
\tautoenunu & { 19.3 $\pm$ 1.1}  \\
\tautomununu  & { 10.8 $\pm$ 0.9} \\
\tautopinu  &  { 19.7 $\pm$ 1.3}  \\
\tautopipiznu   &  {  7.0 $\pm$ 0.5}  \\
 \hline
 Total: & { 9.8 $\pm$ 0.3} \\ 
 \hline
\end{tabular}
  \end{center}
  \label{tab:eff}
\end{table}


In order to determine the  expected number of background events in the data,
we use the final selected data samples with $\eextra$ between  0 and 2.4 \gev.

We first perform an extended unbinned maximum likelihood fit to
the \mes distribution in the final sample,
in the  $\eextra$ sideband  region $ 0.4 \gev < \eextra < 2.4 \gev$.
We use as probability density function(PDF) for the peaking component a Gaussian function joined
to an exponential tail~(Crystal Ball), and as a PDF for the non peaking component a
phase space motivated threshold function~\cite{arguspdf} (Argus).
From this fit, we determine a peaking yield  $N_{pk}^{\rm{high},\rm{data}}$
and signal shape parameters, to be used in later fits.
We apply the same procedure on \BpBm MC events which
pass the final selection and determine the peaking yield $N_{pk}^{\rm{high},\rm{MC}}$.
By fitting \mes  in the $\eextra$ signal region
(with the Crystal Ball parameters fixed to the values determined in the $\eextra$ sideband fits)
in the MC sample, we determine the MC peaking  yield $N_{pk}^{\rm{low},\rm{MC}}$.
We perform a similar fit to data in order to extract the 
amount of combinatorial  background $n_{comb}$,
as the integral of the Argus shaped component in the $\mes > 5.27 \gev$ domain.

We estimate the total background prediction in the signal region as:
\begin{equation}
b = N_{pk}^{\rm{low},\rm{MC}} \times 
\frac{N_{pk}^{\rm{high},\rm{data}}}{N_{pk}^{\rm{high},\rm{MC}}} + n_{comb}
\label{eq:bgformula}
\end{equation}
Due to the low statistics of the final selected sample
an alternative procedure for estimating the background,
which avoids fitting the \mes distribution in the final samples, is also performed.
We fit \mes in the preselection samples
 $\eextra$ sideband to determine the
combinatorial background yield $N_{comb}^{\rm{high},\rm{data\ presel}}$
and in the low $\eextra$ region to determine the
amount of combinatorial background in the preselection signal region
$n^{\rm{presel}}_{comb}$.
 The preselection samples are defined by a looser selection 
only requiring particle identification
for the charged track and a lower bound on the CM momentum 
of 0.6 \gevc for the pion candidate and of
1.0 \gevc for the \pipiz candidate.

We then fit the \mes distribution in the final samples in the $\eextra$ sideband
to determine the combinatorial background yield $N_{comb}^{\rm{high},\rm{data}}$. 
We estimate the combinatorial background in the low
$\eextra$ in the final selected sample $n_{comb}$ scaling
$n^{\rm{presel}}_{comb}$ as
\begin{equation}
  n_{comb} = n^{\rm{presel}}_{comb} \times 
\frac{N_{comb}^{\rm{high},\rm{data}}} {N_{comb}^{\rm{high},\rm{data\ presel}}}
\label{eqn:bgcomb2}
\end{equation}
The same procedure is applied to MC to subtract the 
combinatorial background and determine the
MC peaking yield as:
\begin{equation}
  N_{pk}^{\rm{low},\rm{MC}} = N_{tot}^{\rm{low},\rm{MC}} - n^{\rm{MC\ presel}}_{comb} 
\times \frac{N_{comb}^{\rm{high},\rm{MC}}} {N_{comb}^{\rm{high},\rm{MC\ presel}}}
\label{eqn:bgpeak2}
\end{equation}
This peaking component is scaled by the same scale 
factor ${N_{pk}^{\rm{high},\rm{data}}}/{N_{pk}^{\rm{high},\rm{MC}}}$ of the first method
 and it is added to the combinatorial
of Eq.~(\ref{eqn:bgcomb2}) to obtain the background predictions as in Eq.~(\ref{eq:bgformula}).

We use the mean value between the two methods to obtain 
the final estimate. We determine the error on the estimate 
by adding in quadrature  the statistical error from the 
first method with the half of the difference between the two
values to obtain the total error on the background prediction.
The expected numbers of background events in the four reconstructed modes are shown in Table
\ref{tab:predictions}.

\begin{table}[hbt]
\centering
\caption{
The mean value of expected number of background events
in the signal region, using method A and method B, respectively. The last column shows
the average background prediction 
used in the likelihood scan to extract the signal branching fraction.}
\begin{tabular}{|l|ccc|} \hline
                    &  method A           &  method B           &  average       \\
\hline
\tautoenunu       & $ 1.1 \pm 0.9$      & $ 1.9 \pm 1.0$ & 1.5 $\pm$ 1.4  \\
\tautomununu      & $ 1.9 \pm 1.0$      & $ 1.6 \pm 0.8$ & 1.8  $\pm$ 1.0  \\
\tautopinu        & $ 6.5 \pm 2.0$      & $ 7.1\pm 1.9$  & 6.8  $\pm$ 2.1  \\
\tautopipiznu     & $ 4.1 \pm 1.4$      & $ 4.3\pm 1.3$  & 4.2  $\pm$ 1.4  \\
\hline
All modes         & $ 13.6\pm 2.8$      & $14.9\pm 3.7$  & 14.3 $\pm$  3.0  \\ 
\hline
\end{tabular}
\label{tab:predictions}
\end{table}

\subsection{Results}

\begin{figure}[b]
\centerline{
\setlength{\epsfxsize}{0.99\linewidth}\leavevmode\epsfbox{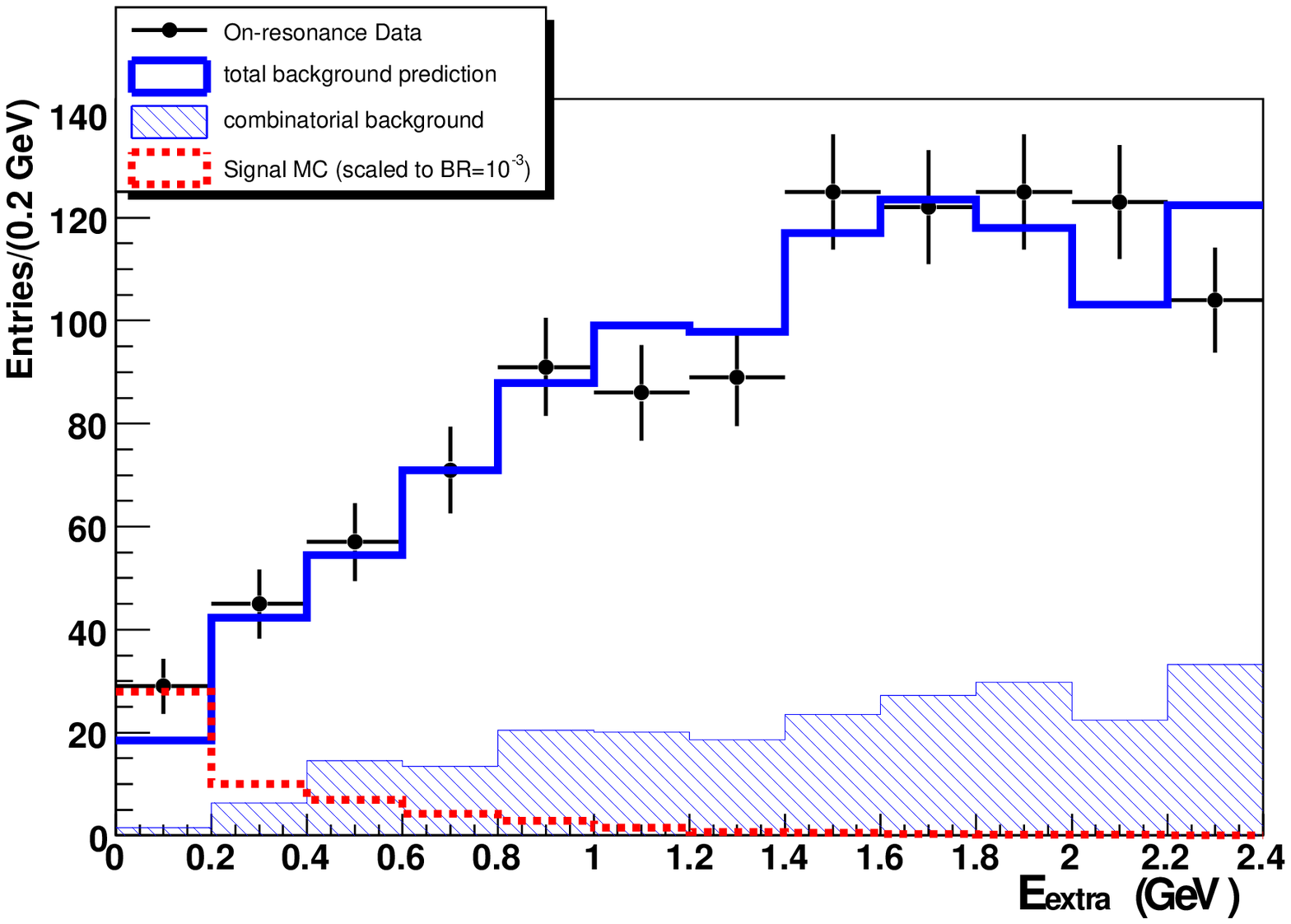}
}
\caption{The $\eextra$ distribution for all the $\tau$ modes combined.}
\label{fig:eextra-all}
\centerline{
\setlength{\epsfxsize}{0.49\linewidth}\leavevmode\epsfbox{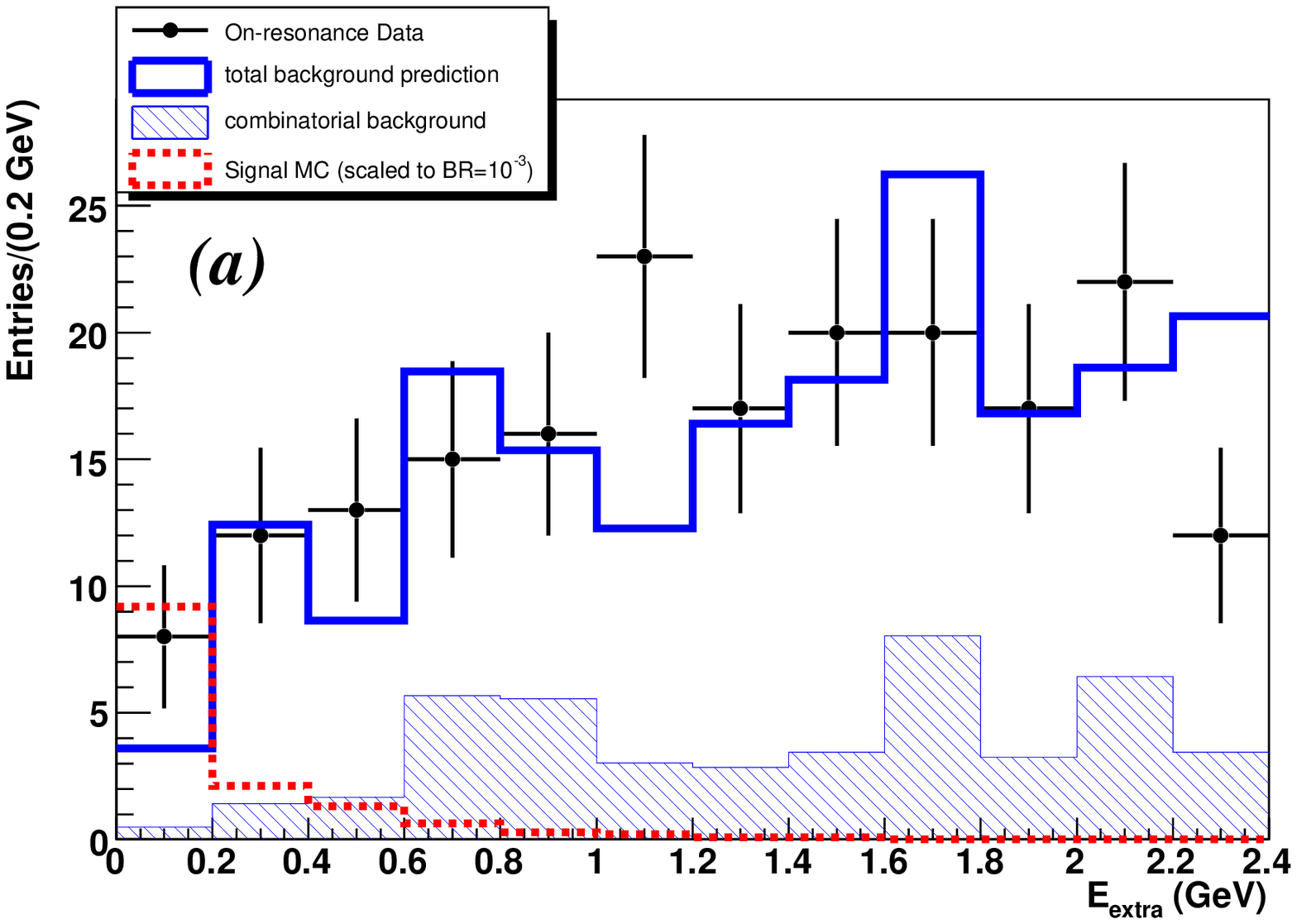}
\setlength{\epsfxsize}{0.49\linewidth}\leavevmode\epsfbox{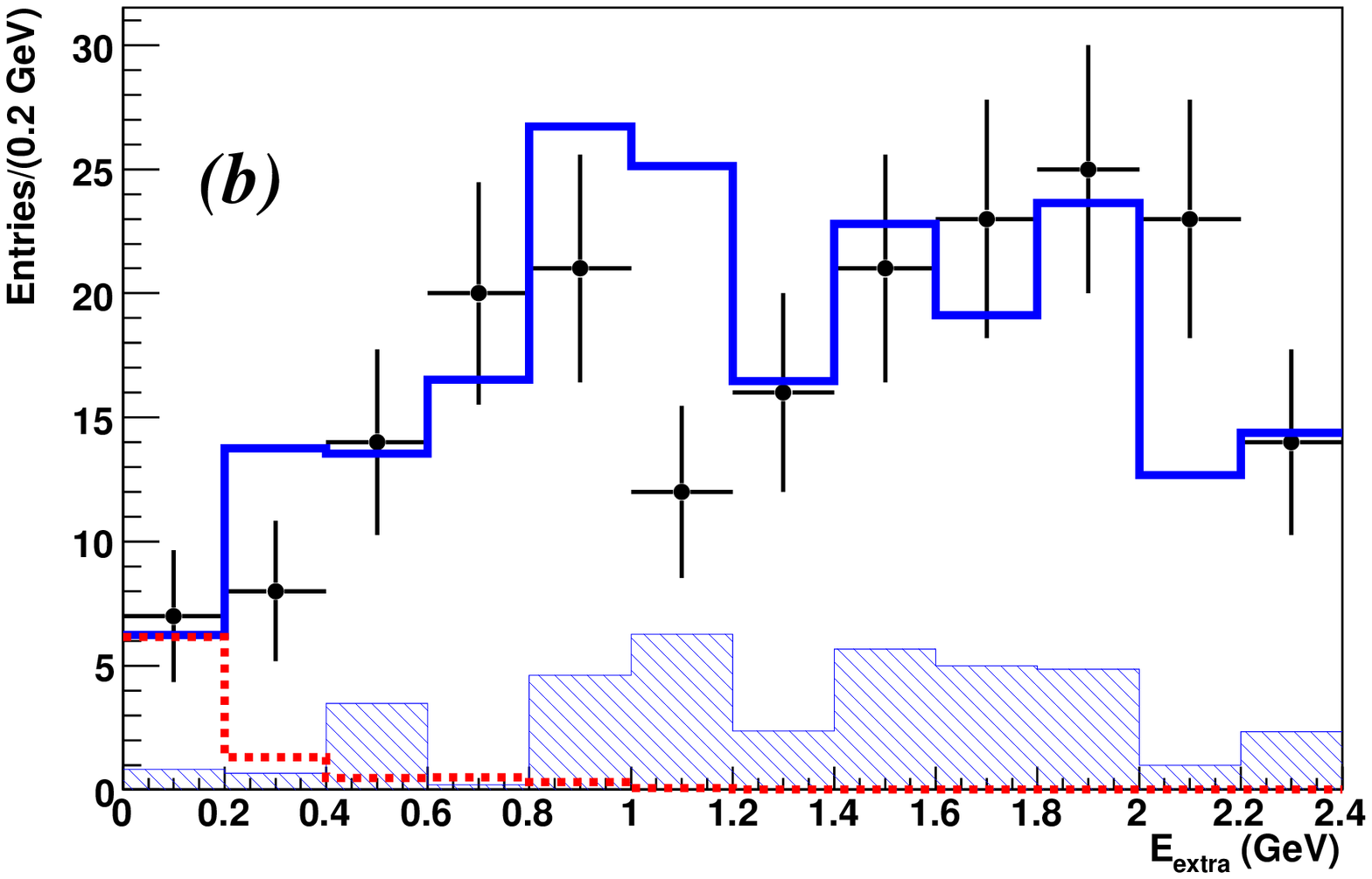}
}
\centerline{
\setlength{\epsfxsize}{0.49\linewidth}\leavevmode\epsfbox{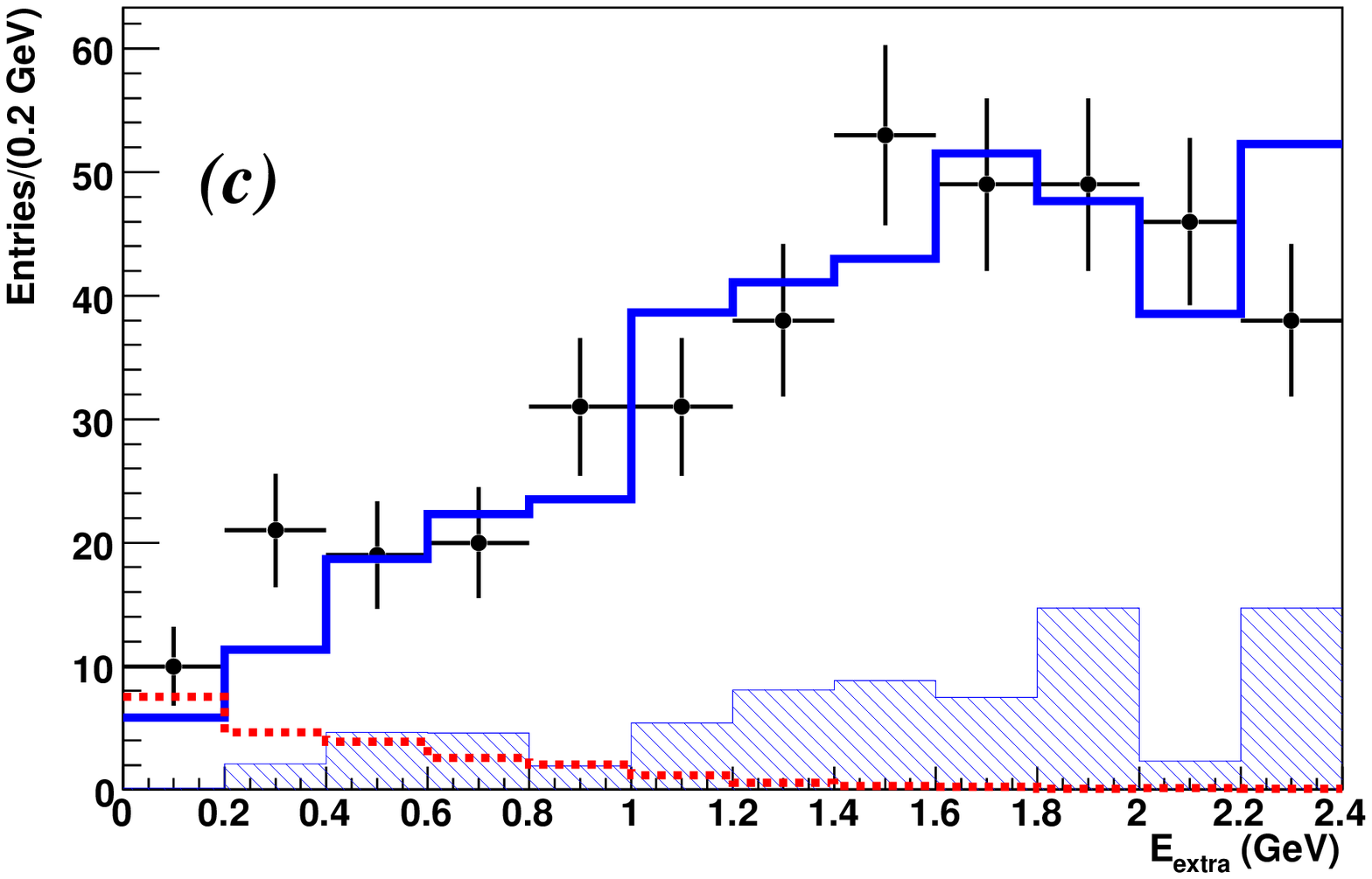}
\setlength{\epsfxsize}{0.49\linewidth}\leavevmode\epsfbox{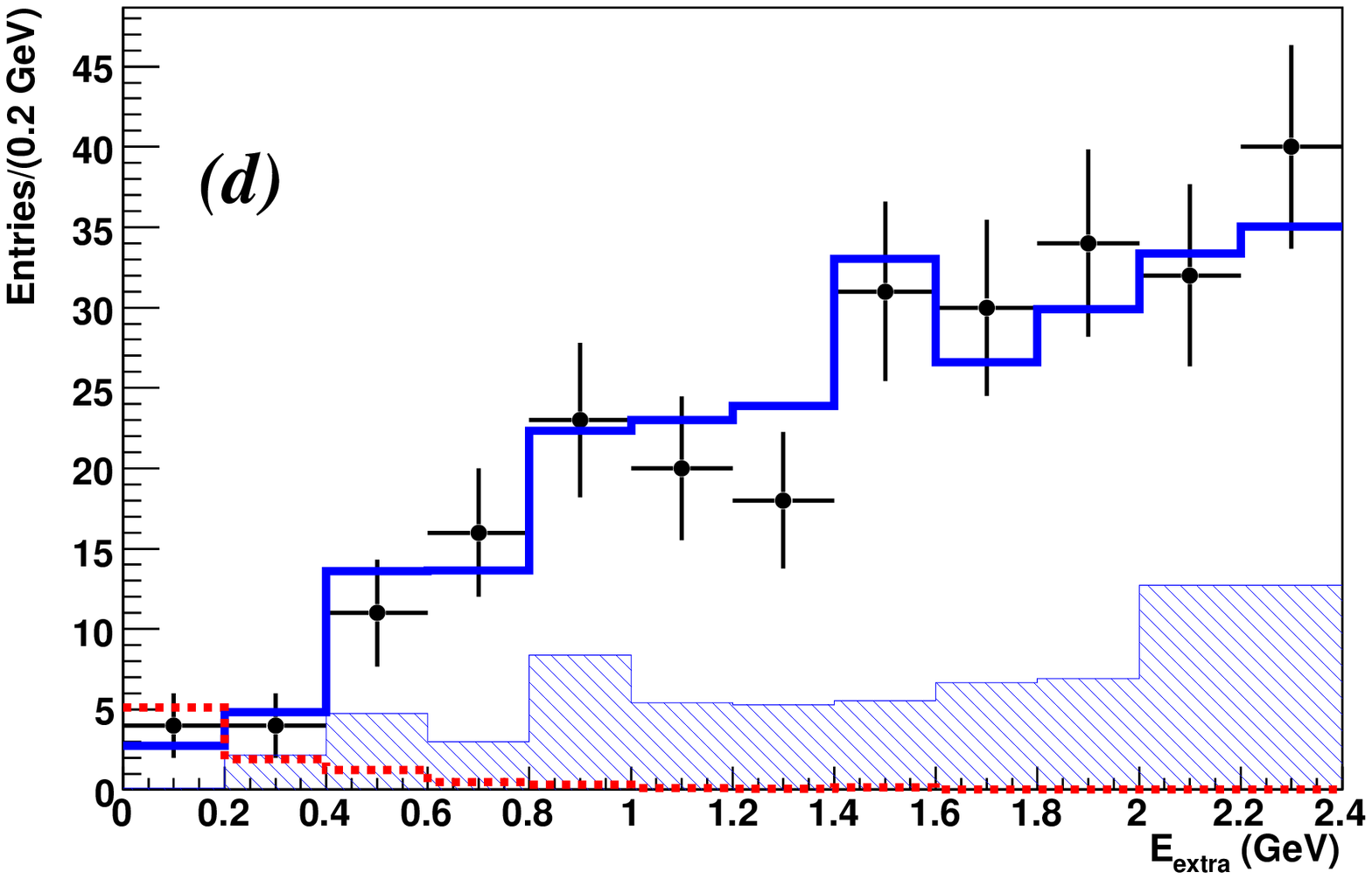}
}
\caption{The $\eextra$ distribution for
the (a) $\tautoenunu$, (b) $\tautomununu$, (c) $\tautopinu$, and (d) $\tautopipiznu$
modes. }
\label{fig:eextra}
\end{figure}

After finalizing the signal selection criteria, we measure the
yield of events in each decay mode in the signal region
of the on-resonance data.
Table~\ref{tab:unblind-result} lists the
number of observed events in on-resonance data in the signal region,
together with the expected number of background events in the
signal region. Figs.~\ref{fig:eextra-all} and \ref{fig:eextra}
show the \eextra distribution for data and
expected background at the end of the selection. The signal MC,
scaled to a branching fraction $\mathcal{B}(\btn) = 10^{-3}$ is overlaid for comparison.
The \eextra distribution is also plotted separately for each $\tau$ decay mode.

\begin{table}[hbt]
\centering
\caption{
The observed number of on-resonance data events in the signal region is
shown, together with the mean number of expected background events.}
\label{tab:unblind-result}
\begin{tabular}{|l|cc|} 
\hline
$\tau$ decay mode   &  background  &  Observed   \\
\hline 
\tautoenunu       & 1.47  $\pm$ 1.37   & 4  \\
\tautomununu      & 1.78  $\pm$ 0.97   & 5  \\
\tautopinu        & 6.79  $\pm$ 2.11   & 10 \\
\tautopipiznu     & 4.23  $\pm$ 1.39   & 5  \\
\hline
All modes    & 14.27 $\pm$  3.03 & 24  \\ 
\hline
\end{tabular}
\end{table}

We determine the \btn branching fraction from the number of fitted signal
candidates $s$ in data according to:
\begin{equation}
\mathcal{B}(\btn) =  \frac { \varepsilon^{\textrm{tag}}_{B} }
 {   \varepsilon^{\textrm{tag}}_{\textrm{sig}} } \times
                \frac { s }{  \varepsilon_{\tau\nu} N^{\textrm{tag}}_{\B^\pm} }
\end{equation}
where $N^{\textrm{tag}}_{\B^\pm}$ is the number of tag $\Bu$ meson correctly reconstructed,
$ \varepsilon^{\textrm{tag}}_{B}$, $\varepsilon^{\textrm{tag}}_{\textrm{sig}}$ 
are the tag $B$ efficiencies in
generic $B\bar{B}$ and signal events respectively,
and $\varepsilon_{ \tau\nu}$ is the efficiency to select 
a signal $B\to\tau\nu$ decay in a tagged event.
The ratio $r_\varepsilon = 
\frac {   \varepsilon^{\textrm{tag}}_{\textrm{sig}} }{ \varepsilon^{\textrm{tag}}_{B} } $
was determined from MC simulation to be $r_\varepsilon = 0.939\pm0.007$.

The results from each of our four signal decay modes ($n_{ch}$)
are combined using the estimator
$Q = {\cal L}(s+b)/{\cal L}(b)$,
where ${\cal L}(s+b)$ and ${\cal L}(b)$ are the
likelihood functions for signal plus background and background-only
hypotheses, respectively:
\begin{equation}
  {\cal L}(s+b) \equiv
  \prod_{i=1}^{n_{ch}}\frac{e^{-(s_i+b_i)}(s_i+b_i)^{n_i}}{n_i!},
        \;
  {\cal L}(b)   \equiv
  \prod_{i=1}^{n_{ch}}\frac{e^{-b_i}b_i^{n_i}}{n_i!}.
  \label{eq:lb}
\end{equation}
\noindent where $s_i = \varepsilon_i s$.
We include the systematic uncertainties, including those of a
statistical nature, on the expected
background ($b_{i}$) in the likelihood definition by
convolving it with a Gaussian function.
The mean of the Gaussian is $b_{i}$, and
the standard deviation ($\sigma_{b_{i}}$) of the Gaussian is the
error on $b_i$~\cite{lista}.

We compute the central value of the branching fraction
(including statistical uncertainty and uncertainty from the 
background) by scanning
over signal branching fraction hypotheses between $0.0$ 
and $3.0 \times 10^{-4}$ in steps
of $0.025\times10^{-4}$ and computing the value of 
$\mathcal{L}(s+b)/\mathcal{L}(b)$ for each
hypothesis. The branching fraction is the hypothesis 
which minimizes 
${Q}= -2 \ln(\mathcal{L}(s+b)/\mathcal{L}(b))$,
and the statistical uncertainty is determined by finding the 
points on the likelihood scan that
occur at one unit above the minimum. 
The systematic error is 
computed for the branching fraction as
a fraction of the central value.


The main sources of uncertainty in the determination of the $\btn$
branching fraction come from the estimation of the tag yield and efficiency,
the reconstruction efficiency of the signal modes and the number of
expected background events.
 We estimate a systematic uncertainty due to the technique used to determine
the tag $B$ yield and reconstruction efficiency, varying the MC based criteria
to define the \BpBm non-peaking component of the \mes shape.
Observing the effect due to the different choices on the branching fraction
measurement, we assign a systematic uncertainty of 3\%.
The systematic uncertainty of the signal efficiencies depends on the
$\tau$ decay mode  and includes the effects of  the tracking of
charged particles, particle identification, and the $\piz$ reconstruction efficiency.
The dominant efficiency uncertainty was found to be due to $E_{extra}$
selection requirement and is found to be 15$\%$.
The contributions due to the tracking and to the choice 
of the $E_{extra}$ selection are treated as correlated
among the different channels.

The systematic uncertainty on the background prediction has been estimated
using two different methods to extract the combinatorial background,
and it has been found to be negligible with respect to the statistical
uncertainty. This uncertainty has been incorporated in the likelihood
definition used to extract the branching fraction.


We determine the branching fraction central value to be
\begin{equation}
\mathcal{B}(\btn) = 1.8^{+1.0}_{-0.9}( \mbox{stat+bkg} )  
\pm 0.3(\mbox{syst})) \times 10^{-4}.
\label{eqn:bf}
\end{equation}
We obtain a significance of 2.7 $\sigma$ from $\Sigma = \sqrt{\textrm{Q}_{\textrm{min}}}$,
where $\textrm{Q}_{\textrm{min}}$ is the minimum value of the likelihood ratio, 
if we do not include the
uncertainties on the background predictions.
If we include the background uncertainty into account we obtain the smaller significance of
2.2~$\sigma$.
We compute the 90\% C.L. upper limit using the $CL_s$ method ~\cite{cls}
to be $ \mathcal{B}(\btn)  < 3.4 \times 10^{-4}$.


\subsection{Combined Results}

\begin{figure}[!thb]
  \begin{center}
    \includegraphics[width=0.99\linewidth]{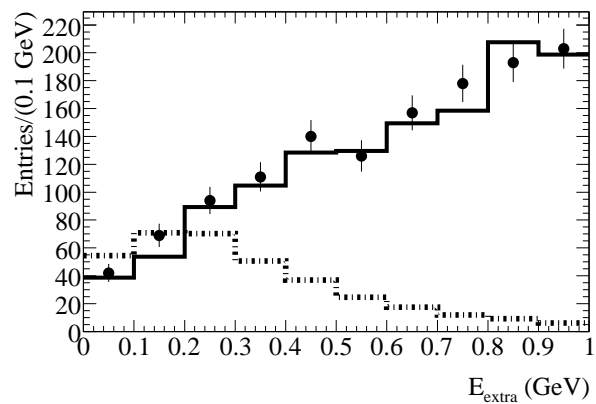}
    \caption{
The $\eextra$ distribution for all the $\tau$ modes combined
in the analysis with semileptonic tag.
The dashed line shows predicted signal distribution assuming
the $B^{\pm} \to \tau^{\pm} \nu$  branching fraction of $10^{-3}$.
}
    \label{fig:semileptonic}
  \end{center}
\end{figure}

We report here the branching fraction obtained by 
combining the hadronic tag analysis result,
described in this document, with the $\babar$ semileptonic 
tag analysis result, based on a statistically
independent data sample and reported in ~\cite{babar-note1654}.
Both the analyses proceed in a similar manner for the
four $\tau$ decay mode reconstruction and
use the same likelihood ratio scan technique 
to combine the different $\tau$ decay modes.
Fig.~\ref{fig:semileptonic}
shows the $\eextra$ distribution for all the $\tau$ modes combined
in the analysis with semileptonic tag.

The likelihood ratio is extended to combine eight reconstruction modes, 
four from the hadronic tag and four from the semileptonic tag.
We obtain
\begin{eqnarray}
\mathcal{B}(\btn) = ~~~~~~~~~~~~~~~~~~~~~~~~~~~~~~~~~~~~~\cr 
( 1.20^{+0.40}_{-0.38}( \mbox{stat} ) ^{+0.29}_{-0.30} (\mbox{bkg})  
\pm 0.22(\mbox{syst})) \times 10^{-4},
\label{eqn:bfcombined}
\end{eqnarray}
where the first error is statistical, the second is due 
to the background prediction systematic uncertainty,
and the third one is due to the other systematic sources.
We obtain a significance of 2.6 $\sigma$ including the uncertainty on the expected
background~(3.2 $\sigma$ if this uncertainty is not included).

\begin{figure}[!thb]
  \begin{center}
    \includegraphics[width=0.99\linewidth]{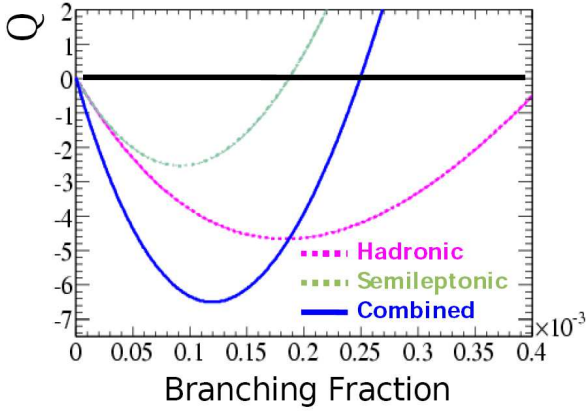}
    \caption{
Scan of $Q=-2ln( \mathcal{L}(s+b)/\mathcal{L}(b) )$ as a function
of the $B^{\pm} \to \tau^{\pm} \nu$ decay branching fraction. 
The uncertainty on the expected number of background events
has been included in the likelihood definition.
The results are shown for the hadronic tag, semileptonic tag, and
combined analysis.
}
    \label{fig:logqerr}
  \end{center}
\end{figure}

Fig.~\ref{fig:logqerr} shows the $\mathcal{L}(s+b)/\mathcal{L}(b)$ 
scan as a function
of signal branching fraction (the background uncertainty is included
in the likelihood).

\begin{figure}[!t]
\centerline{
\setlength{\epsfxsize}{0.99\linewidth}\leavevmode\epsfbox{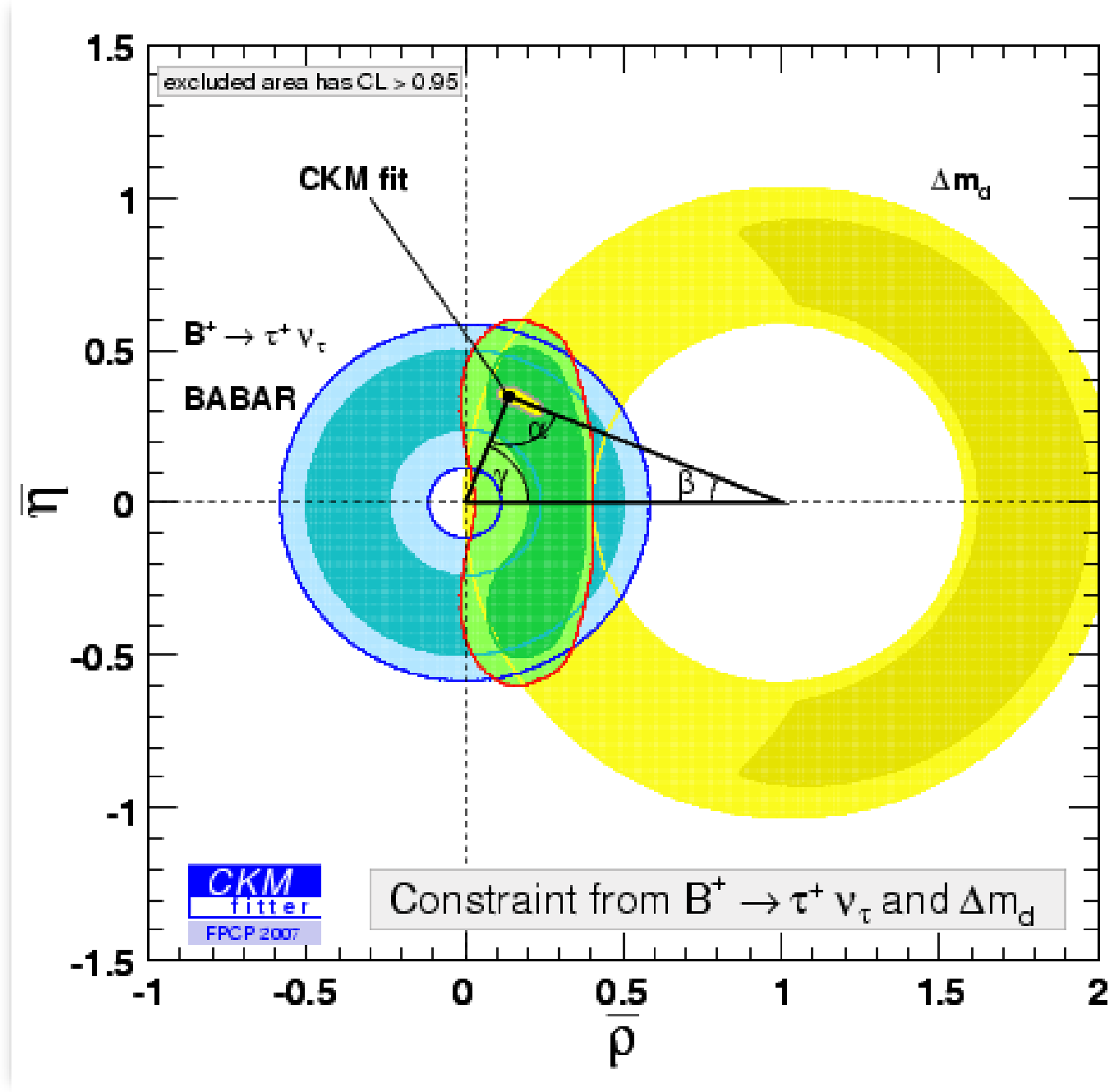}
}
\caption{Constraints at 95$\%$ C.L. on $\rho$ and $\eta$ parameters 
(through measurement of $V_{ub}$) of the
CKM triangle using  $B^{\pm} \to \tau^{\pm} \nu$ results presented here,
see Eq.~(\ref{eqn:br}).
Constraints from $B_d$-mixing are also shown, along with combined results~\cite{ckm}.
}
\label{fig:taunu-ckm}
\centerline{
\setlength{\epsfxsize}{0.99\linewidth}\leavevmode\epsfbox{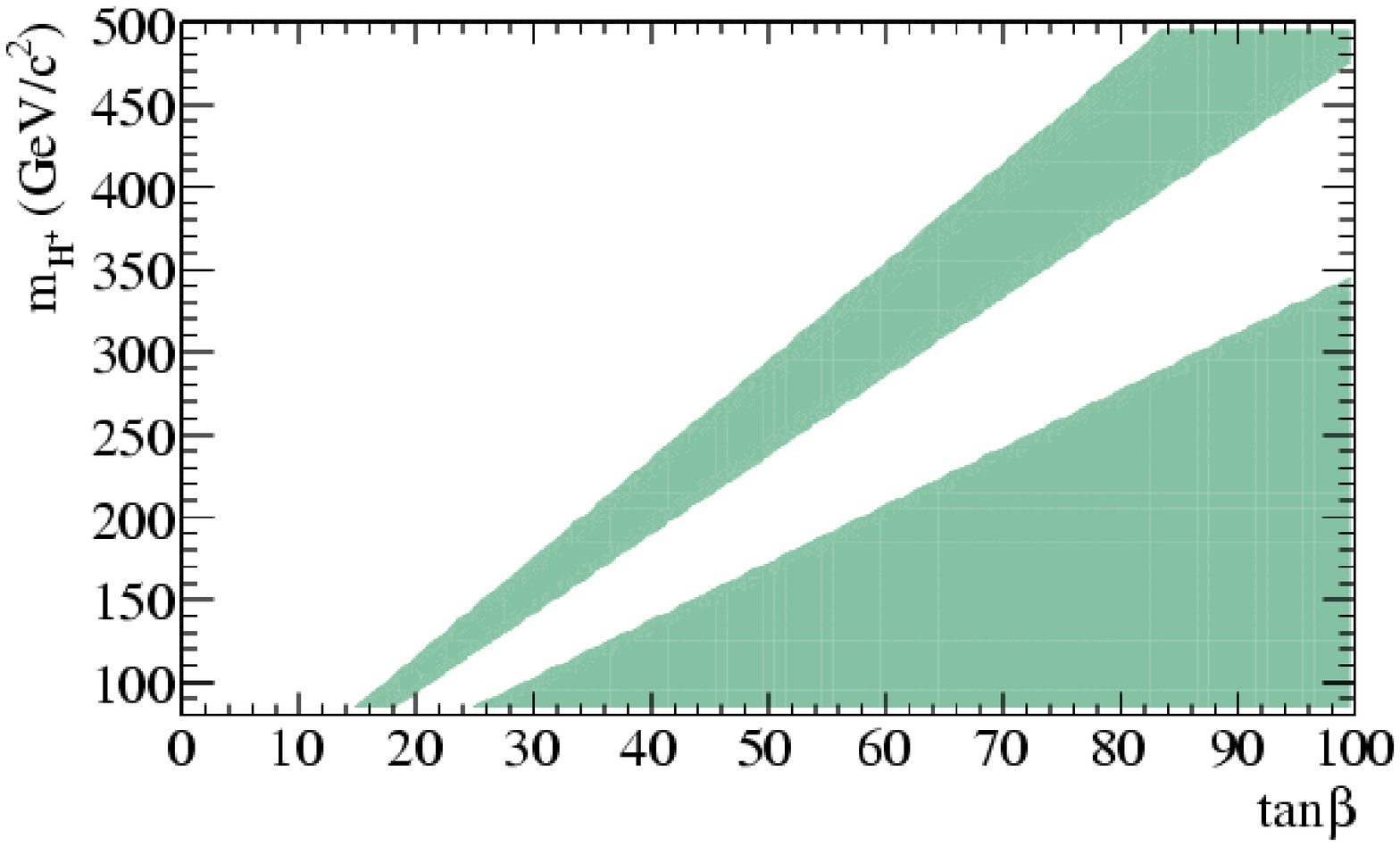}
}
\caption{Constraints on $\tan\beta$ and charged Higgs mass
from $B^{\pm} \to \tau^{\pm} \nu$ results presented here, 
see Eqs.~(\ref{eqn:br}) and (\ref{eqn:brmod}).
Shaded areas are excluded at 95$\%$ C.L. 
}
\label{fig:taunu-higgsconst}
\end{figure}

\subsection{Discussion}

In summary, we report here preliminary branching fraction 
of the $B^{\pm} \to \tau^{\pm} \nu$ decay using two independent
methods of tag $B$ reconstruction: hadronic and semileptonic
$B$ decay in the recoil.
We find a 2.6 $\sigma$ (3.2 $\sigma$ not including expected background 
uncertainty) excess in data which can be converted to a 
preliminary branching fraction central value of 
$\mathcal{B}(B^{\pm} \to \tau^{\pm} \nu)=
({1.20}^{+0.40+0.29}_{-0.38-0.30}\pm0.22)\times{10^{-4}}$.

Given the measurement of the $B^{\pm} \to \tau^{\pm} \nu$ branching
fraction, we may constraint CKM parameter $V_{ub}$
using Eq.~(\ref{eqn:br}). See Fig~\ref{fig:taunu-ckm} for illustration.
Alternatively, given the best world measurements of parameters 
present in Eq.~(\ref{eqn:br}), we can constraint 
$\tan\beta$ and charged Higgs mass in the model with two Higgs doublets,
as shown in Fig~\ref{fig:taunu-higgsconst}.



\section{\boldmath $B^\pm\to\varphi(1020)K^{*}(892)^\pm$}

\begin{figure}[t]
\centerline{
\setlength{\epsfxsize}{0.99\linewidth}\leavevmode\epsfbox{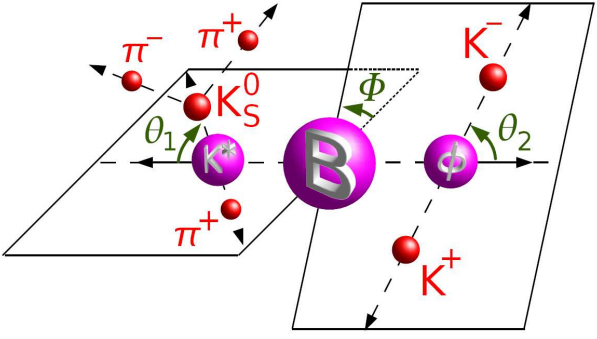}
}
\vspace{-0.3cm}
\caption{
Definition of decay angles given in the rest frames of the decaying parents.
\label{fig:helicity}
}
\end{figure}

One of our goals is to test the quark-spin projections in the
$B\to\varphi K^{*}$ decay, as shown 
in Fig.~\ref{fig:feyn-phikst}.
However, equivalently we can measure
spin projections of the two vector mesons in the $B$ decay.
There are three
complex amplitudes for the three spin projections possible,
and therefore there are a total of 12 non-trivial independent
real parameters describing the $B$ and $\Bbar$ decays.
Information about the amplitude sizes and phases can be deduced
from analysis of angular distribution as illustrated in 
Fig.~\ref{fig:helicity}.



\subsection{Analysis Method}

The $B^\pm\to\varphi(1020)K^{*\pm}\to(K^+K^-)(K\pi)^\pm$ candidates
are analyzed with two $(K\pi)^\pm$ final states, $K^0_S\pi^\pm$ and $K^\pm\pi^0$.
The neutral pseudoscalar mesons are reconstructed
in the final states $K^0_S\to\pi^+\pi^-$ and $\pi^0\to\gamma\gamma$.
We define the helicity angle $\theta_i$ as the angle between the direction
of the $K$ or $K^+$ meson from $K^*\to K\pi$ ($\theta_1$) or $\varphi\to K^+K^-$
($\theta_2$) and the direction opposite the $B$ in the $K^*$ or $\varphi$
rest frame, and $\Phi$ as the angle between the decay planes of the two
systems, see Fig.~\ref{fig:helicity}.
The differential decay width has four complex amplitudes
$A_{J\lambda}$ which describe two spin states of the $K\pi$ system
($J=1$ or $0$) and the three helicity states of the $J=1$ state
($\lambda=0$ or $\pm 1$):
\begin{eqnarray}
\label{eq:helicityfull}
{d^3\Gamma \over d{\cal H}_1 d{\cal H}_2d\Phi} \propto
\left|~\sum_{}
A_{J\lambda} Y_{J}^{\lambda}({\cal H}_1,\Phi) Y_{1}^{-\!\lambda}(-{\cal H}_2,0)~\right|^2,
\end{eqnarray}
where ${\cal H}_i=\cos\theta_i$ and
$Y_{J}^{\lambda}$ are the spherical harmonics with $J=1$
for $K^{*}(892)$ and $J=0$ for $(K\pi)_0^{*}$.
We reparameterize the amplitudes as
$A_{1\pm1}=(A_{1\parallel}\pm A_{1\perp})/\sqrt{2}$.

We identify $B$ meson candidates using two kinematic variables:
$m_{\rm{ES}} = [{ (s/2 + \mathbf{p}_{\Upsilon} \cdot
\mathbf{p}_B)^2 / E_{\Upsilon}^2 - \mathbf{p}_B^{\,2} }]^{1/2}$
and $\Delta{E}=(E_{\Upsilon}E_B-\mathbf{p}_{\Upsilon}\cdot\mathbf{p}_B-s/2)/\sqrt{s}$,
where $(E_B,\mathbf{p}_B)$ is the four-momentum of the $B$ candidate,
and $(E_{\Upsilon},\mathbf{p}_{\Upsilon})$ is the $e^+e^-$ initial state four-momentum,
both in the laboratory frame.
We require $m_{\rm{ES}}>5.25$ GeV and $|\Delta{E}|<0.1$ GeV.
The requirements on the invariant masses are
$0.75 < m_{K\!\pi} < 1.05$ GeV, $0.99 < m_{K\!\Kbar} < 1.05$ GeV,
$|m_{\pi\pi}-m_{K^0}|< 12$ MeV, and $120 < m_{\gamma\gamma} < 150$ MeV
for the $K^{*\pm}$, $\varphi$, $K^0_S$, and $\pi^0$, respectively.
For the $K^0_S$ candidates, we also require the cosine of the angle
between the flight direction from the interaction point
and momentum direction to be greater than 0.995
and the measured proper decay time greater than five times its
uncertainty.

To reject the dominant $e^+e^-\to$ quark-antiquark
background, we use the angle $\theta_T$ between the $B$-candidate
thrust axis and that of the rest of the event, and a Fisher
discriminant ${\cal F}$~\cite{bigPRD}. Both variables are
calculated in the center-of-mass frame. The discriminant
combines the polar angles of the $B$-momentum vector
and the $B$-candidate thrust axis with respect to the beam axis,
and two moments of the energy flow around the
$B$-candidate thrust axis~\cite{bigPRD}.

To reduce combinatorial background with low-momentum
$\pi^0$ candidates, we require ${\cal H}_1<0.6$.
When more than one candidate is reconstructed,
which happens in $7\%$ of events with $K_S^0$ and $17\%$ with $\pi^0$,
we select the one whose $\chi^2$ of the charged-track vertex fit
combined with $\chi^2$ of the invariant mass consistency of
the $K^0_S$ or $\pi^0$ candidate, is the lowest.
We define the $b$-quark flavor sign $Q$ to be opposite
to the charge of the $B$ meson candidate.


We use an unbinned, extended maximum-likelihood fit~\cite{babar:vv, babar:vt}
to extract the event yields $n_{j}^k$ and the parameters of the probability
density function (PDF) ${\cal P}_{j}^k$.
The index $j$ represents three event categories used in our data model:
the signal $B^\pm\to\varphi(K\pi)^\pm$ ($j = 1$),
a possible background from $B^\pm\to f_0(980)K^{*\pm}$ ($j = 2$),
and combinatorial background ($j = 3$).
The superscript $k$ corresponds to the value of $Q=\pm$
and allows for a $C\!P$-violating difference between
the $B^+$ and $B^-$ decay amplitudes ($A$ and $\Abar$).
In the signal category, the yield and asymmetry of the
$B^\pm\to\varphi K^*(892)^\pm$ mode, $n_{\rm sig}$ and ${\cal A}_{C\!P}$,
and those of the $B^\pm\to\varphi(K\pi)_0^{*\pm}$ mode are parameterized
by applying the fraction of $\varphi K^*(892)^\pm$ yield, $\mu^k$, to $n_1^k$.
Hence, $n_{\rm sig}=n_1^+\times\mu^+ + n_1^-\times\mu^-$,
${\cal A}_{C\!P}=(n_1^+\times\mu^+ - n_1^-\times\mu^-)/n_{\rm sig}$,
and the $\varphi(K\pi)_0^{*\pm}$ yield is $n_1^+\times(1-\mu^+) + n_1^-\times(1-\mu^-)$.

The likelihood ${\cal L}_i$ for each candidate $i$ is defined as
${\cal L}_i = \sum_{j,k}n_{j}^k\,
{\cal P}_{j}^k$({\boldmath ${\rm x}_i$};~$\mu^k$,~{\boldmath$\zeta$},~{\boldmath$\xi$}),
where the PDF is formed based on the following set of observables
{\boldmath ${\rm x}_i$}~$=\{{\cal H}_1$, ${\cal H}_2$, $\Phi$,
$m_{K\!\pi}$, $m_{K\!\Kbar}$, $\Delta E$, $m_{\rm{ES}}$, ${\cal F}$, $Q$\}
and the dependence on $\mu^k$ and polarization parameters {\boldmath$\zeta$}
is relevant only for the signal PDF ${\cal P}_{1}^k$.
The remaining PDF parameters {\boldmath$\xi$}
are left free to vary in the fit for the combinatorial
background and are fixed to the values extracted from
Monte Carlo (MC) simulation~\cite{geant} and calibration
$B\to\Dbar\pi$ decays for event categories $j = 1$ and $2$.

The helicity part of the signal PDF is the
ideal angular distribution from Eq.~(\ref{eq:helicityfull}),
multiplied by an empirical acceptance function
${\cal{G}}({\cal H}_1,{\cal H}_2,\Phi)
\equiv{\cal{G}}_1({\cal H}_1)\times{\cal{G}}_2({\cal H}_2)$.
Here, the amplitudes $A_{J\lambda}$ are expressed in terms of
the polarization parameters
{\boldmath$\zeta$}~$\equiv\{f_L$, $f_{\perp}$, $\phi_{\parallel}$,
$\phi_{\perp}$, $\delta_0$, ${\cal A}_{C\!P}^0$, ${\cal A}_{C\!P}^{\perp}$,
$\Delta \phi_{\parallel}$, $\Delta \phi_{\perp}$, $\Delta\delta_0$\}
defined in Table~\ref{tab:results}.
$C\!P$-violating differences are incorporated via the replacements
in Eq.~(\ref{eq:helicityfull}) for $B^{+}$ decays:
$f_L\to f_L\times(1+{\cal A}^0_{C\!P}\times Q)$,
$f_\perp\to f_\perp\times(1+{\cal A}^\perp_{C\!P}\times Q)$,
$\phi_\parallel\to(\phi_\parallel+\Delta\phi_\parallel\times Q)$,
$\phi_\perp\to(\phi_\perp+{\pi/2}+(\Delta\phi_\perp+{\pi/2})\times Q)$,
and $\delta_0\to(\delta_0+\Delta\delta_0\times Q)$.

A relativistic spin-$J$ Breit--Wigner amplitude parameterization
is used for the resonance masses~\cite{pdg2006,f0mass}, and the
$(K\pi)^{*\pm}_0$ $m_{K\!\pi}$ amplitude is parameterized with
the LASS function~\cite{Aston:1987ir}.
The latter includes the $K_0^{*}(1430)^\pm$ resonance
together with a nonresonant component.
The interference between the $J=0$ and $1$ $(K\pi)^{\pm}$ contributions
is modeled with the three terms $2{\cal R\rm e}(A_{1\lambda} A^*_{00})$
in Eq.~(\ref{eq:helicityfull}) with the four-dimensional angular and $m_{K\!\pi}$
parameterization and with dependence on $\mu^k$ and {\boldmath$\zeta$}.

\begin{figure}[t]
\centerline{
\setlength{\epsfxsize}{0.5\linewidth}\leavevmode\epsfbox{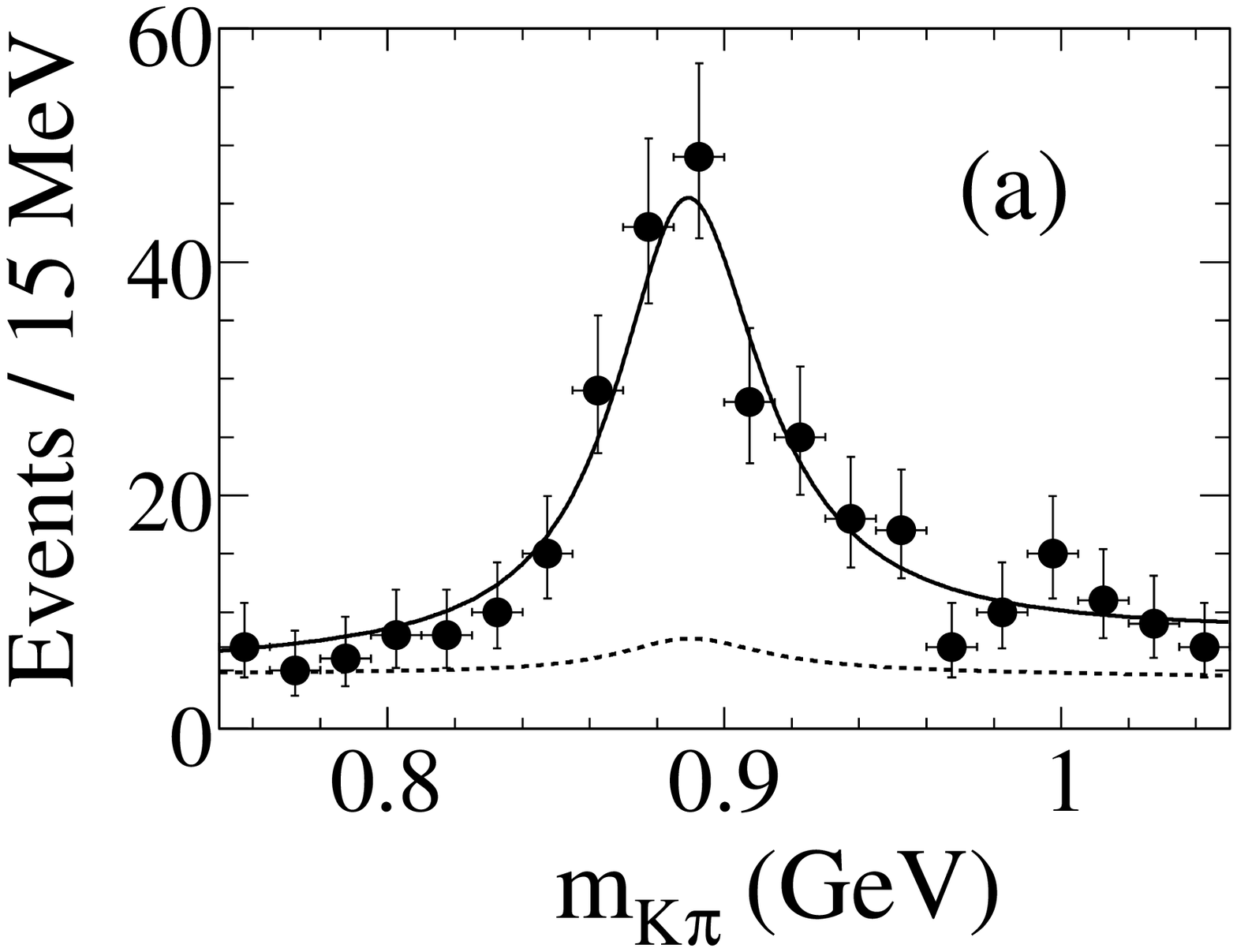}
\setlength{\epsfxsize}{0.5\linewidth}\leavevmode\epsfbox{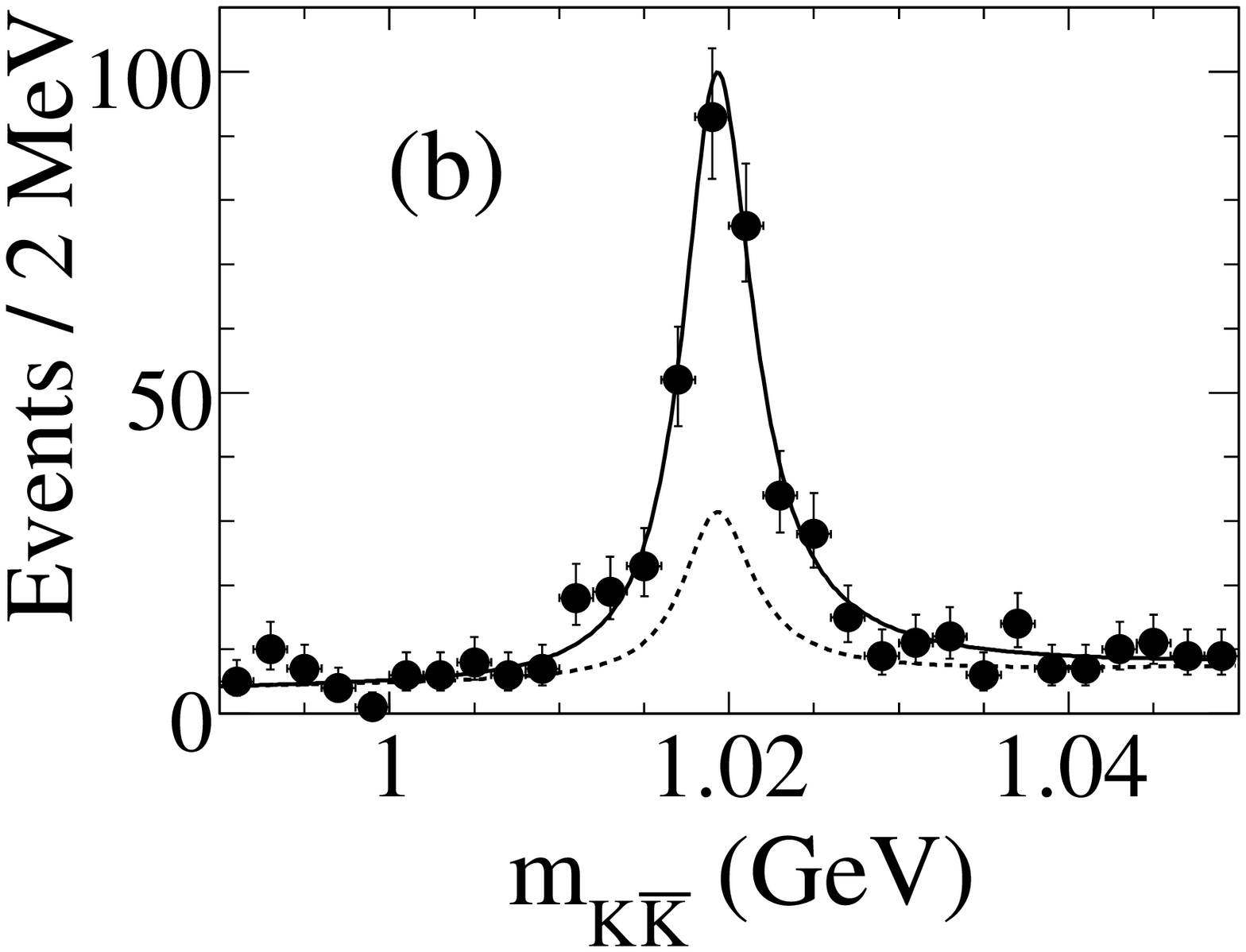}
}
\centerline{
\setlength{\epsfxsize}{0.5\linewidth}\leavevmode\epsfbox{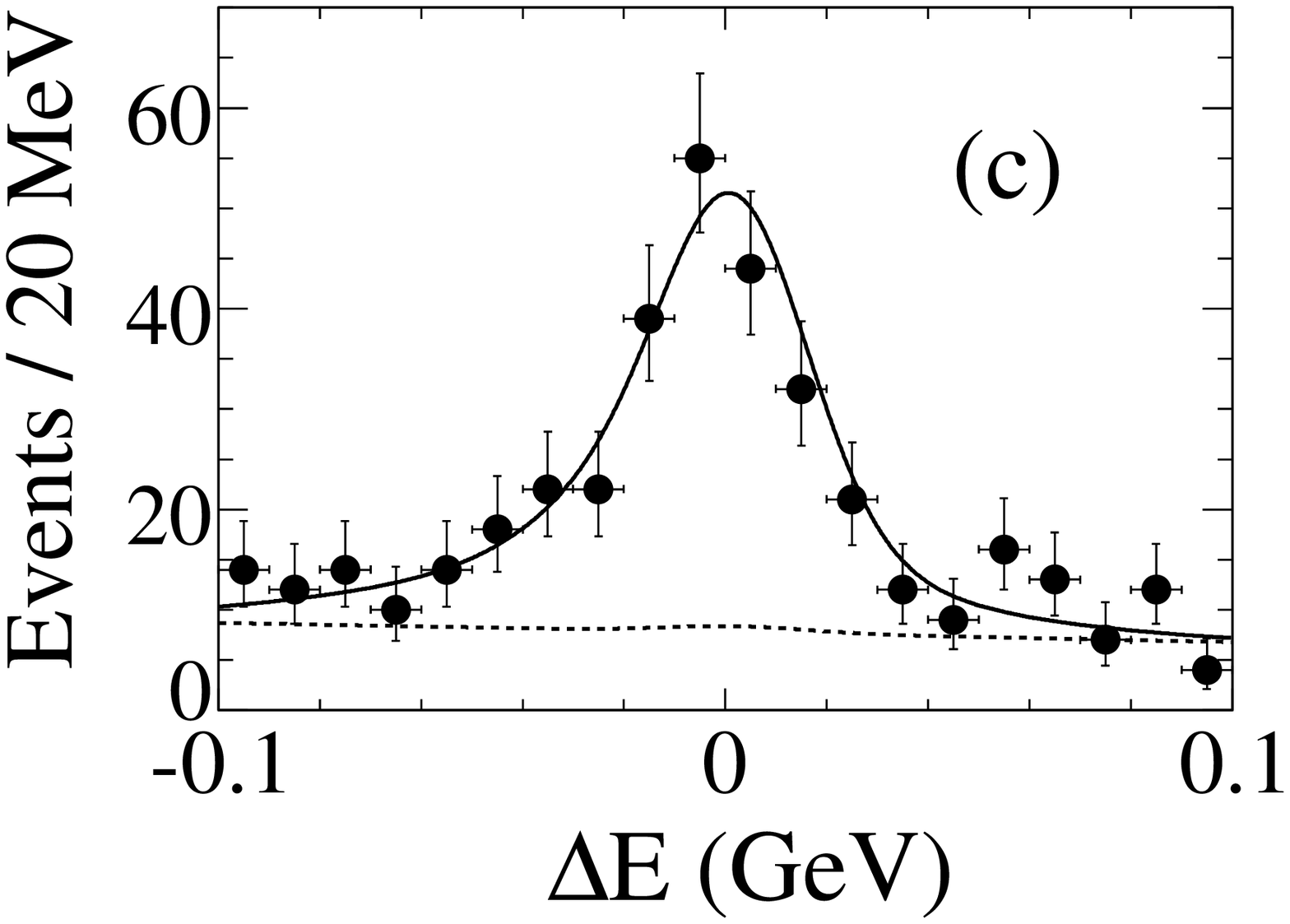}
\setlength{\epsfxsize}{0.5\linewidth}\leavevmode\epsfbox{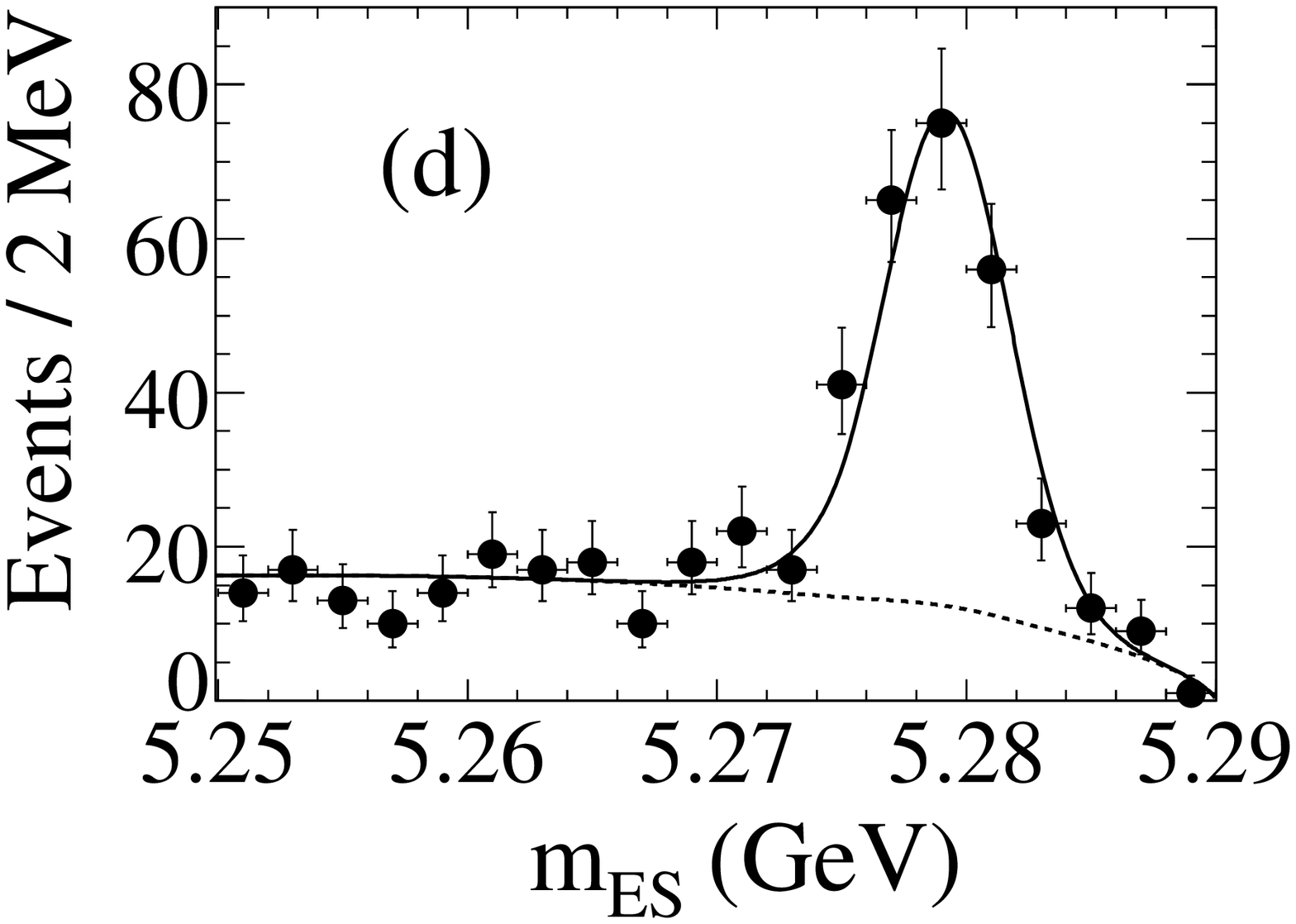}
}
\vspace{-0.3cm}
\caption{\label{fig:projection1}
Projections onto the variables (a) $m_{K\!\pi}$, (b) $m_{K\!\Kbar}$, (c) $\Delta E$,
and (d) $m_{\rm ES}$ for the signal $B^\pm\to\varphi(K\pi)^\pm$ candidates
with a requirement discussed in the text.
The solid (dashed) lines show the signal-plus-background
(background) PDF projections.
}
\end{figure}
\begin{figure}[t]
\centerline{
\setlength{\epsfxsize}{0.5\linewidth}\leavevmode\epsfbox{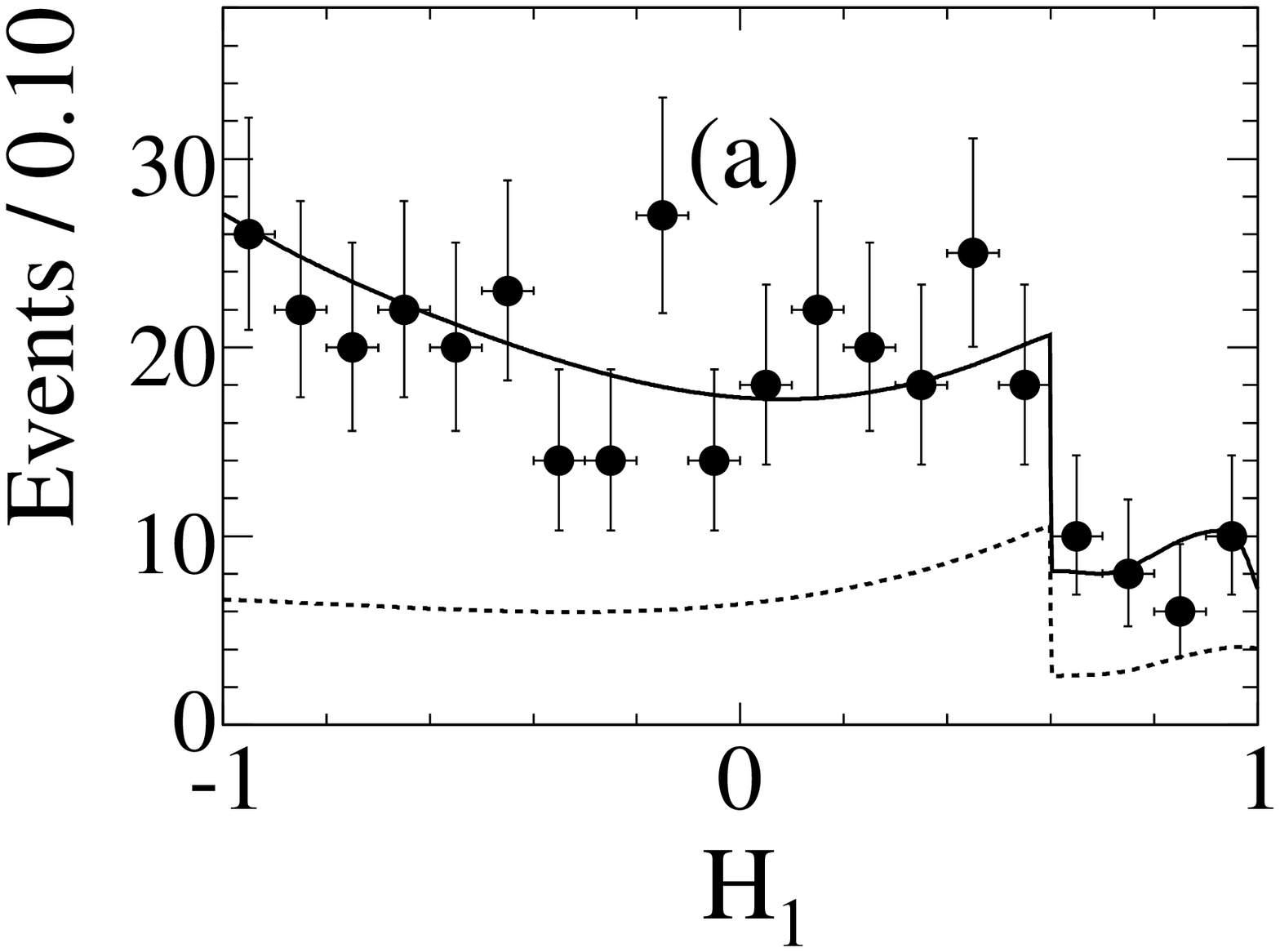}
\setlength{\epsfxsize}{0.5\linewidth}\leavevmode\epsfbox{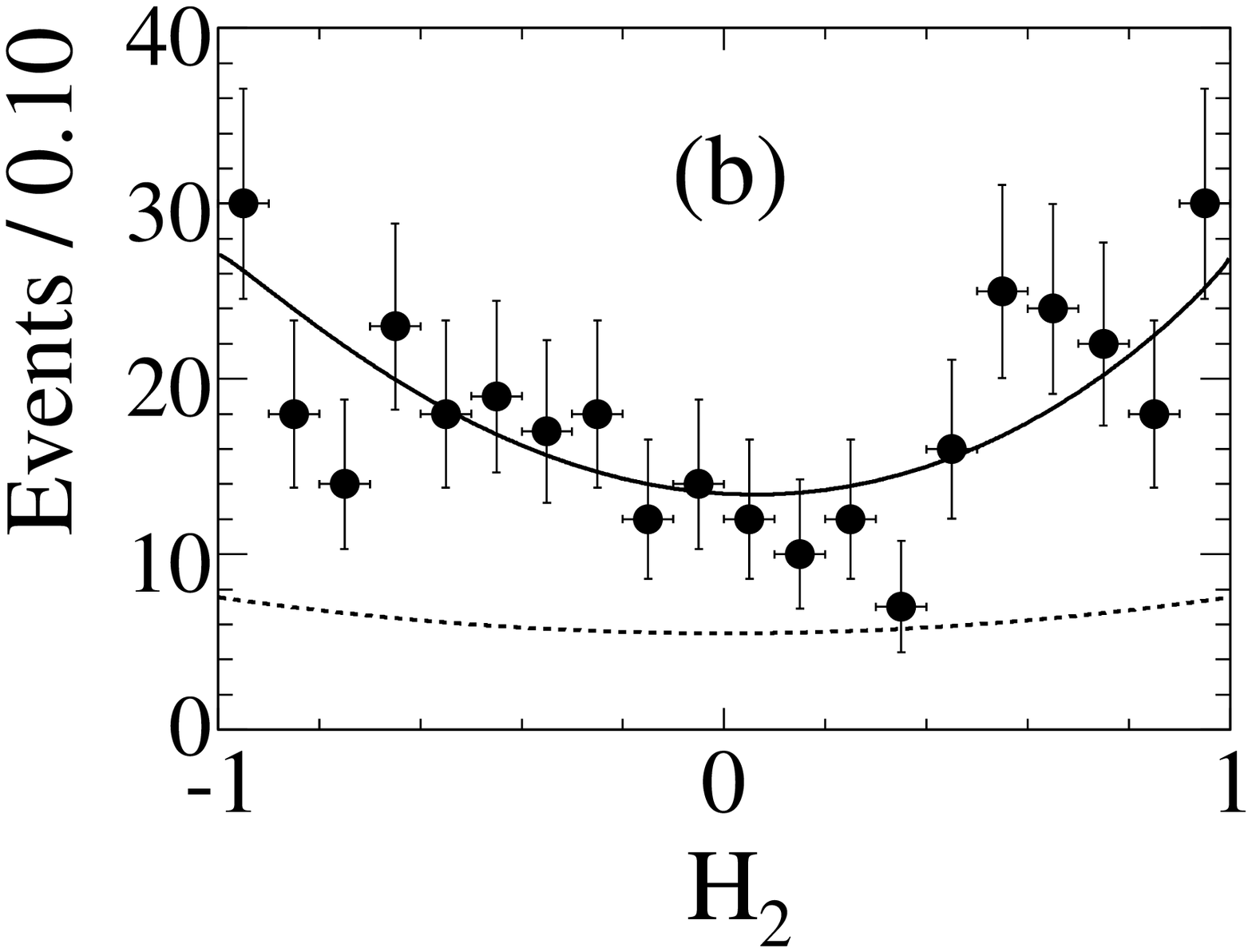}
}
\centerline{
\setlength{\epsfxsize}{0.5\linewidth}\leavevmode\epsfbox{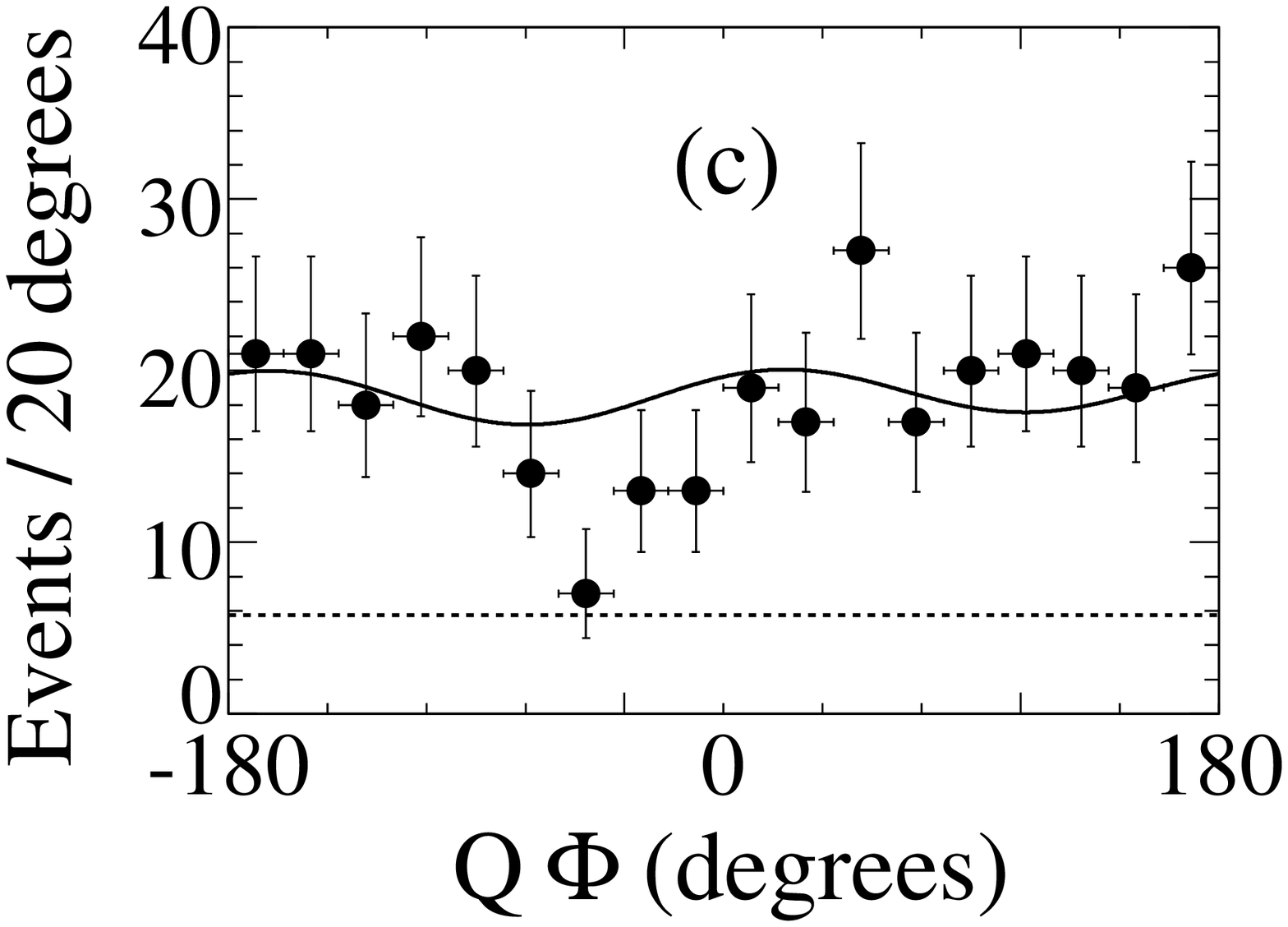}
\setlength{\epsfxsize}{0.5\linewidth}\leavevmode\epsfbox{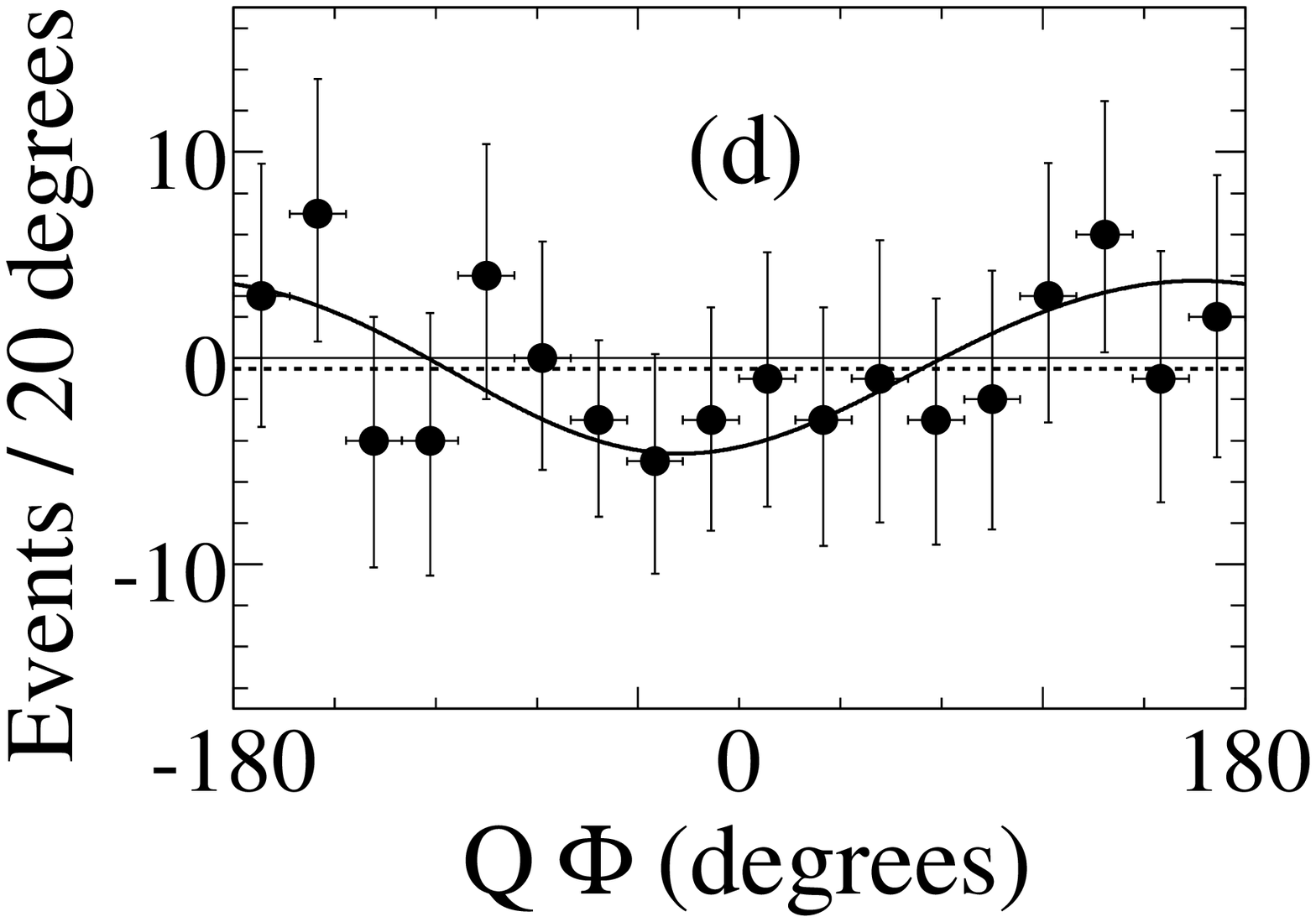}
}
\vspace{-0.3cm}
\caption{\label{fig:projection2}
Projections onto the variables (a)
${\cal H}_1$, (b) ${\cal H}_2$, (c) $Q~\!\Phi$, and (d)
the differences between
the $Q~\!\Phi$ projections for events with
${\cal H}_1~\!{\cal H}_2>0$ and with
${\cal H}_1~\!{\cal H}_2<0$
for the signal $B^\pm\to\varphi(K\pi)^\pm$  candidates
following the solid (dashed) line definitions in Fig.~\ref{fig:projection1}.
The step in the ${\cal H}_1$ PDF distributions is due to the
selection requirement  ${\cal H}_1<0.6$ in the $B^\pm\to\varphi(K^\pm\pi^0)$
channel.
}
\end{figure}

\begingroup
\begin{table*}[t]
\caption{\label{tab:results}
Summary of results for the $B^\pm\to\varphi K^{*}(892)^\pm$ decay.
The twelve primary results are presented for the two decay
subchannels along with the combined results, where the branching
fraction ${\cal B}$ is computed using the number of signal events
$n_{\rm sig}$ and the total selection efficiency $\varepsilon$,
which includes the daughter branching fractions~\cite{pdg2006} and
the reconstruction efficiency $\varepsilon_{\rm reco}$ obtained
from MC simulation.
The definition of the six $C\!P$-violating parameters allows for
differences between the $B^+$ and ${B}^-$ decay amplitudes
$A$ and $\Abar$ with superscript $Q=-$ and $+$, respectively.
The systematic uncertainties are quoted last and are not included
for the intermediate primary results in each subchannel.
The dominant fit correlation coefficients (${\cal C}$) are presented, where
we show correlations of ${\delta_0}$  with ${\phi_\parallel}/{\phi_\perp}$
and of ${\Delta\delta_0}$ with ${\Delta\phi_\parallel}/{\Delta\phi_\perp}$.
}
\begin{center}
{
\begin{tabular}{cccccc}
\hline
\vspace{-3mm} & & \\
   parameter
 & definition
 & $K^{*}(892)^\pm\to K^0_S\pi^\pm$
 & $K^{*}(892)^\pm\to K^\pm\pi^0$
 & combined
 & ${\cal C}$
\cr
\vspace{-3mm} & & & \\
\hline
\vspace{-3mm} & & & \\
  ${\cal B}$ 
 & $\Gamma/\Gamma_{\rm total}$
 & $(10.5\pm1.4)\times10^{-6}$
 & $(11.6\pm1.5)\times10^{-6}$
 & $(11.2\pm1.0\pm0.9)\times10^{-6}$
\cr
\vspace{-3mm} & & & \\
  ${f_L}$  
 & ${|A_{10}|^2/\Sigma|A_{1\lambda}|^2}$
 & $0.51\pm0.07$
 & $0.46^{+0.10}_{-0.09}$
 & $0.49\pm0.05\pm 0.03$
 & \multirow{2}{13mm}{~{\Large\}}$-58\%$}
\cr
\vspace{-3mm} & & & \\
  ${f_\perp}$ 
 & ${|A_{1\perp}|^2/\Sigma|A_{1\lambda}|^2}$
 & $0.22^{+0.07}_{-0.06}$
 & $0.21^{+0.09}_{-0.08}$
 & $0.21\pm 0.05\pm 0.02$
 &
\cr
\vspace{-3mm} & & & \\
  ${\phi_\parallel}-\pi$ 
 & ${\rm arg}(A_{1\parallel}/A_{10})-\pi$
 & $-0.75^{+0.28}_{-0.24}$
 & $-0.77\pm0.35$
 & $-0.67\pm{0.20}\pm0.07$
 &  \multirow{2}{13mm}{~{\Large\}}~$+56\%$}
\cr
\vspace{-3mm} & & & \\
  ${\phi_\perp}-\pi$  
 & ${\rm arg}(A_{1\perp}/A_{10})-\pi$
 & $-0.15\pm 0.24$
 & $-0.89^{+0.40}_{-0.46}$
 & $-0.45\pm{0.20}\pm0.03$
 &
\cr
\vspace{-3mm} & & & \\
  ${\delta_0}-\pi$  
 & ${\rm arg}(A_{00}/A_{10})-\pi$
 & $-0.25\pm0.24$
 & $+0.11\pm{0.31}$
 & $-0.07\pm0.18\pm0.06$
 &                        {$+37\%/+36\%$}
\cr
\vspace{0mm} & & & \\
  ${\cal A}_{C\!P}$ 
 & $(\Gamma^+-\Gamma^-)/(\Gamma^++\Gamma^-)$
 & $-0.09\pm0.13$
 & $+0.07\pm0.13$
 & $0.00\pm0.09\pm0.04$
 &
\cr
\vspace{-3mm} & & & \\
  ${\cal A}_{C\!P}^0$ 
 & $(f_L^+-f_L^-)/(f_L^++f_L^-)$
 & $+0.24\pm0.15$
 & $+0.09\pm0.20$
 & $+0.17\pm0.11\pm0.02$
 &  \multirow{2}{13mm}{~{\Large\}}$-50\%$}
\cr
\vspace{-3mm} & & & \\
  ${\cal A}_{C\!P}^{\perp}$ 
 & $(f_\perp^+-f_\perp^-)/(f_\perp^++f_\perp^-)$
 & $+0.12\pm{0.31}$
 & $+0.41^{+0.54}_{-0.40}$
 & $+0.22\pm{0.24}\pm0.08$
 &
\cr
\vspace{-3mm} & & & \\
  $\Delta \phi_{\parallel}$  
 & $(\phi_{\parallel}^+-\phi_{\parallel}^-)/2$
 & $+0.02\pm0.28$
 & $+0.22\pm{0.35}$
 & $+0.07\pm0.20\pm0.05$
 &  \multirow{2}{13mm}{~{\Large\}}~$+57\%$}
\cr
\vspace{-3mm} & & & \\
  $\Delta \phi_{\perp}$  
 & $(\phi_{\perp}^+-\phi_{\perp}^--\pi)/2$
 & $+0.18\pm 0.24$
 & $+0.48^{+0.46}_{-0.40}$
 & $+0.19\pm{0.20}\pm0.07$
 &
\cr
\vspace{-3mm} & & & \\
  ${\Delta\delta_0}$  
 & $(\delta_0^+-\delta_0^-)/2$
 & $+0.13\pm{0.24}$
 & $+0.34\pm0.31$
 & $+0.20\pm0.18\pm0.03$
 &                        {$+37\%/+37\%$}
\cr
\vspace{0mm} & & & \\
   $n_{\rm sig}$ 
 &
 & $102\pm 13\pm 6$
 & $117^{+15}_{-16}\pm 7$
 &
 &
\cr
\vspace{-3mm} & & & \\
    $\varepsilon$  
 &
 & $(2.53\pm0.13)$ \%
 & $(2.59\pm0.17)$ \%
 &
 &
\cr
\vspace{-3mm} & & & \\
   $\varepsilon_{\rm reco}$  
 &
 & $(22.3\pm1.2)$ \%
 & $(16.0\pm1.0)$ \%
 &
 &
\cr
\vspace{-3mm} & & & \\
\hline
\end{tabular}
}
\end{center}
\end{table*}
\endgroup

\begin{figure}[t]
\centerline{
\setlength{\epsfxsize}{0.99\linewidth}\leavevmode\epsfbox{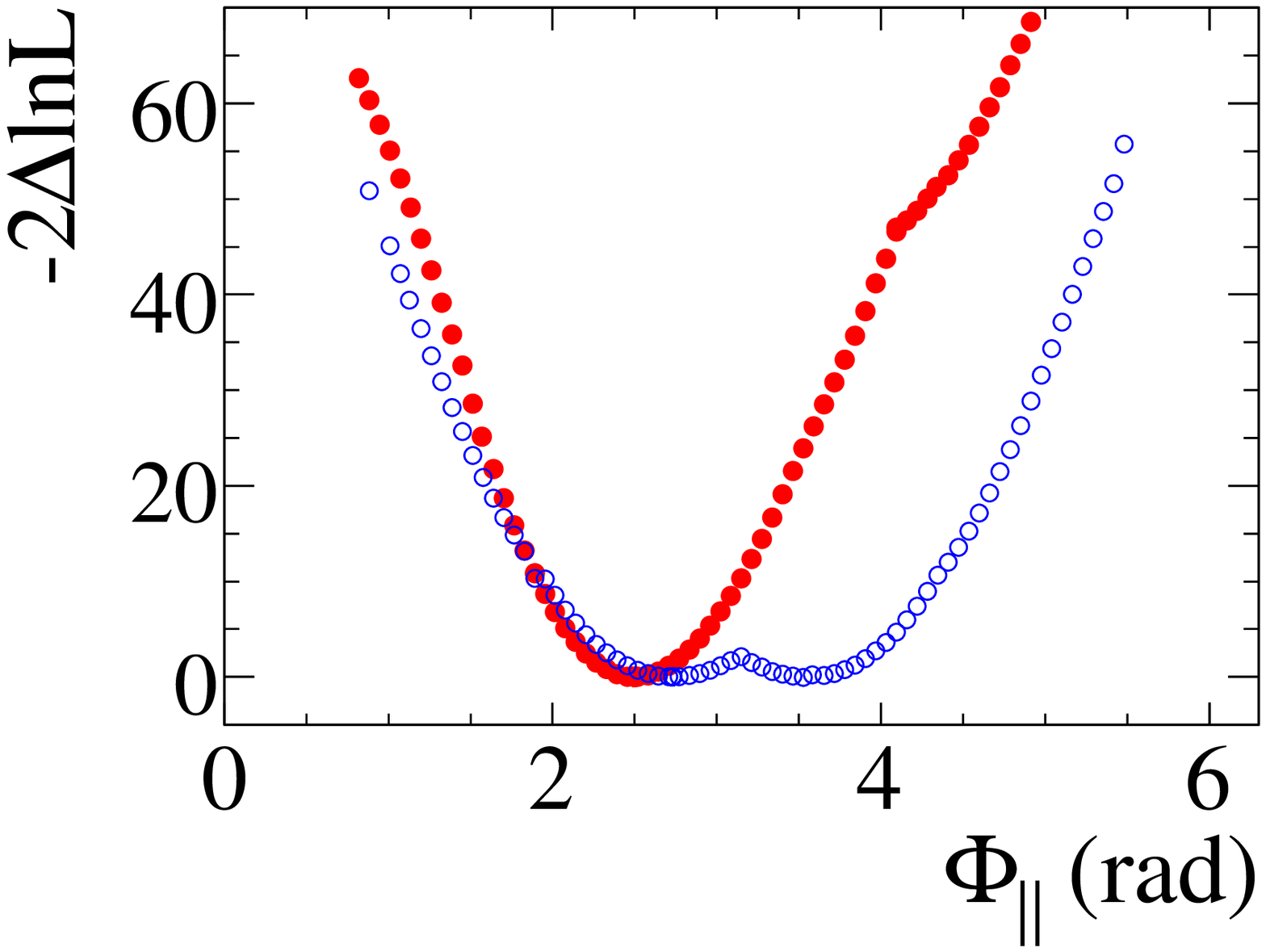}
}
\vspace{-0.3cm}
\caption{
Distribution of $2\ln{\cal L}$ as a function of $\phi_\parallel$
for the two fit configurations: including P- and S-wave $K\pi$ 
interference into account (solid points) and ignoring it (open points).
The former resolves the phase ambiguity.
\label{fig:philscann}
}
\centerline{
\setlength{\epsfxsize}{0.99\linewidth}\leavevmode\epsfbox{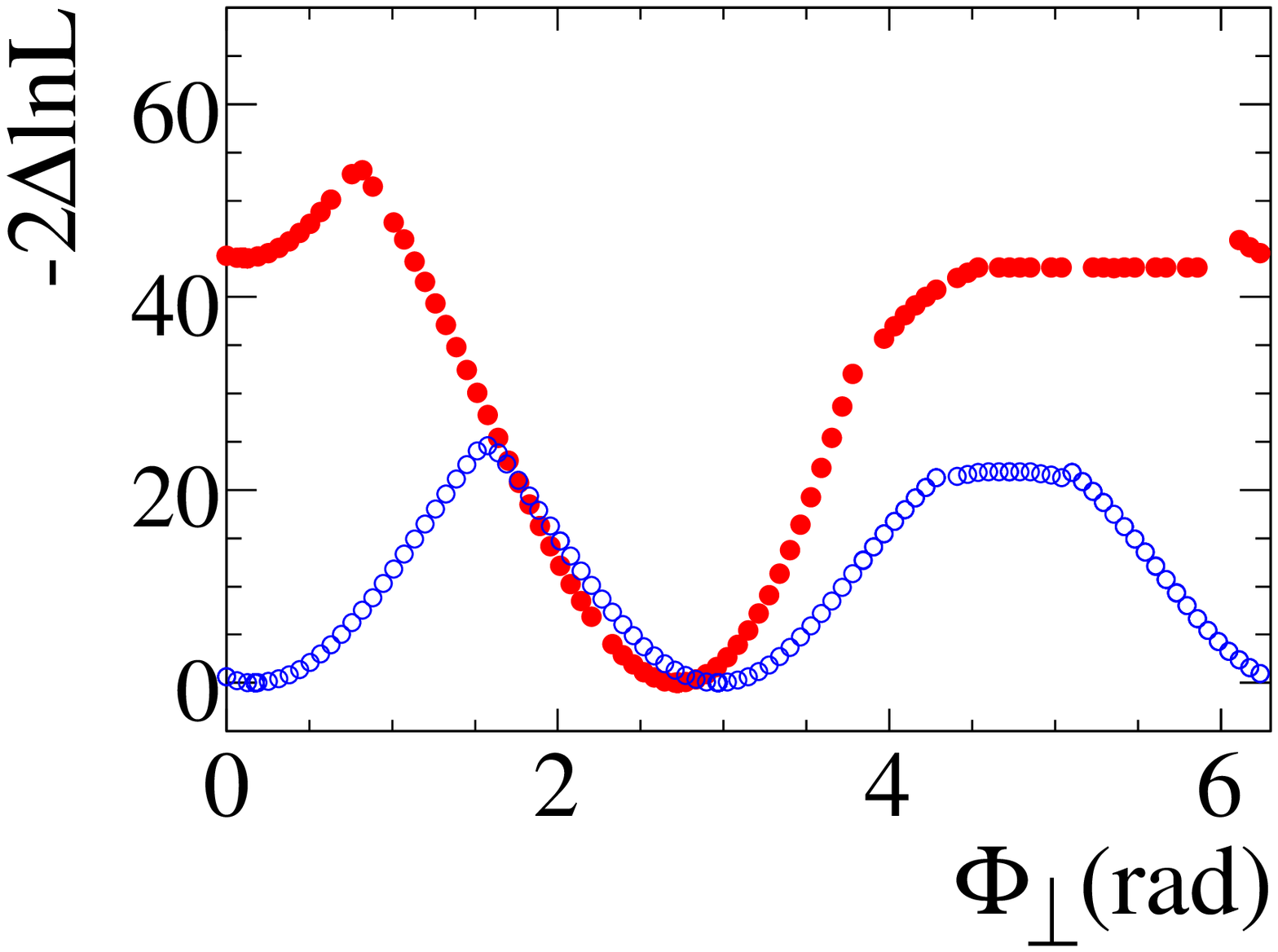}
}
\vspace{-0.3cm}
\caption{
Distribution of $2\ln{\cal L}$ as a function of $\phi_\perp$
for the two fit configurations: including P- and S-wave $K\pi$ 
interference into account (solid points) and ignoring it (open points).
The former resolves the phase ambiguity.
\label{fig:phidscann}
}
\end{figure}

The signal PDF
for a given candidate $i$ is a joint PDF for the helicity angles
and resonance mass as discussed above, and the product of
the PDFs for each of the remaining variables.
The combinatorial background PDF is the product of the
PDFs for independent variables and is found to describe
well both the dominant quark-antiquark background and the
background from random combinations of $B$ tracks.
The signal and background PDFs are illustrated in
Figs.~\ref{fig:projection1} and~\ref{fig:projection2}.
For illustration,
the signal fraction is enhanced with a requirement on the
signal-to-background probability ratio, calculated with
the plotted variable excluded, that is at least 50\%
efficient for signal $B^\pm\to\varphi(K\pi)^\pm$ events.
We use a sum of Gaussian functions
for the parameterization of the signal PDFs
for $\Delta E$, $m_{\rm{ES}}$, and ${\cal F}$.
For the combinatorial background, we use polynomials,
except for $m_{\rm{ES}}$ and ${\cal F}$ distributions
which are parameterized by an empirical phase-space
function and by Gaussian functions, respectively.
Resonance production occurs in the background and
is taken into account in the PDF.


\subsection{Results}

We observe a nonzero $B^\pm\to\varphi K^{*}(892)^\pm$
yield with significance, including systematic uncertainties,
of more than 10$\sigma$.
The significance is defined as the square root of the change in
$2\ln{\cal L}$ when the yield is constrained to zero in the
likelihood ${\cal L}$.
In Table~\ref{tab:results}, results of the fit are presented,
where the combined results are obtained from the simultaneous fit
to the two decay subchannels.


We repeat the fit by varying the fixed parameters
in {\boldmath$\xi$} within their uncertainties
and obtain the associated systematic uncertainties.
We allow for a flavor-dependent acceptance function
and reconstruction efficiency in the study of asymmetries.
The biases from the finite resolution of the angle measurements,
the dilution due to the presence of fake combinations,
or other imperfections in the signal PDF model are estimated
with MC simulation.

\begingroup
\begin{table*}[t]
\caption{\label{tab:results-all}
Summary of branching fraction (${\cal B}$) and 
longitudinal polarization fraction (${f_L}$) results 
for the various $B\to\varphi K^{*}$ decay on $\babar$.
The quantum numbers $J^P$ are given for the $K^{*}$ resonances.
For a complete list of up to 12 independent parameters measured
in the $\varphi K^{*}(892)^0$ and $\varphi K_2^{*}(1430)^0$ decay modes
see Ref.~\cite{babar:vt}.
}
\begin{center}
{
\begin{tabular}{|c|c|c|c|}
\hline
~~$J^P$~~ & ~~$B$ decay mode~~ & ${\cal B}$ranching (${ 10^{-6}}$) & ${f_L}$ \cr
\hline
\vspace{0mm} {$0^+$} & {$\varphi K^{*}_0(1430)^0$} & $4.6\pm{0.7}\pm 0.6$ &  \cr
\vspace{0mm} {$1^-$} & {$\varphi K^{*}(892)^0$} & ${9.2}\pm{0.7}\pm 0.6$ & ${0.51}\pm{0.04}\pm 0.02$ \cr
\vspace{0mm} {$1^-$} &{ $\varphi K^{*}(892)^+$} & ${11.2}\pm1.0\pm0.9$ & ${0.49}\pm0.05\pm 0.03$ \cr
\vspace{0mm} $1^-$ & $\varphi K^{*}(1680)^0$ & $<{3.5}~~(0.7^{+1.0}_{-0.7}\pm 1.1)$ & -- \cr
\vspace{0mm} {$2^+$} & {$\varphi K_2^{*}(1430)^0$} & ${7.8}\pm{1.1}\pm 0.6$ & ${0.85}^{+0.06}_{-0.07}\pm 0.04$ \cr
\vspace{0mm} $3^-$ & $\varphi K_3^{*}(1780)^0$ &  $<{2.7}~(-0.9\pm{1.4}\pm 1.1)$ & --\cr
\vspace{0mm} $4^+$ & $\varphi K_4^{*}(2045)^0$ &  $<{15.3}~~(6.0^{+4.8}_{-4.0}\pm 4.1)$ & -- \cr
\hline
\end{tabular}
}
\end{center}
\end{table*}
\endgroup

The nonresonant $K^+K^-$ contribution under the
$\varphi$ is accounted for with the $B^0\to f_0 K^{*0}$ category.
Its yield is consistent with zero.
The $m_{K\!\Kbar}$ PDF shape in this category
is varied from the resonant to phase-space
and the yield is varied from the observed value to the extrapolation from the
neutral $B$-decay mode~\cite{babar:vt} to estimate the systematic uncertainties.
Additional systematic uncertainty originates
from other potential $B$ backgrounds,
which we estimate can contribute
at most a few events to the signal component.
The systematic uncertainties in efficiencies are dominated
by those in particle identification, track finding,
and $K^0_S$ and $\pi^0$ selection.
Other systematic effects arise from event-selection criteria,
$\varphi$ and $K^{*0}$ branching fractions, and the number of $B$ mesons.


\begin{figure}[b]
\centerline{\setlength{\epsfxsize}{1.0\linewidth}\leavevmode\epsfbox{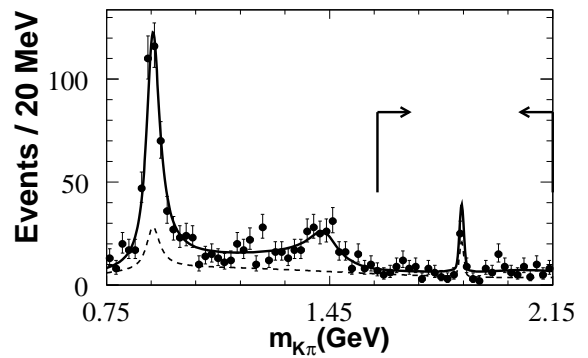}}
\caption{\label{fig:kpimass-neutral}
Distribution of the $K\pi$ invariant mass extended above 1.6 GeV
from the study of $B^0\to\phi(K^+\pi^-)$ decays in Ref.~\cite{babar:vt}.
The data distribution is shown with a requirement
to enhance the signal as discussed in regard
to Fig.~\ref{fig:projection1} in text.
The solid (dashed) line shows the signal-plus-background
(background) expected distributions.
The arrows indicate the higher mass range,
1.60 to 2.15 GeV.
}
\end{figure}

\begin{figure}[t]
\centerline{
\setlength{\epsfxsize}{0.99\linewidth}\leavevmode\epsfbox{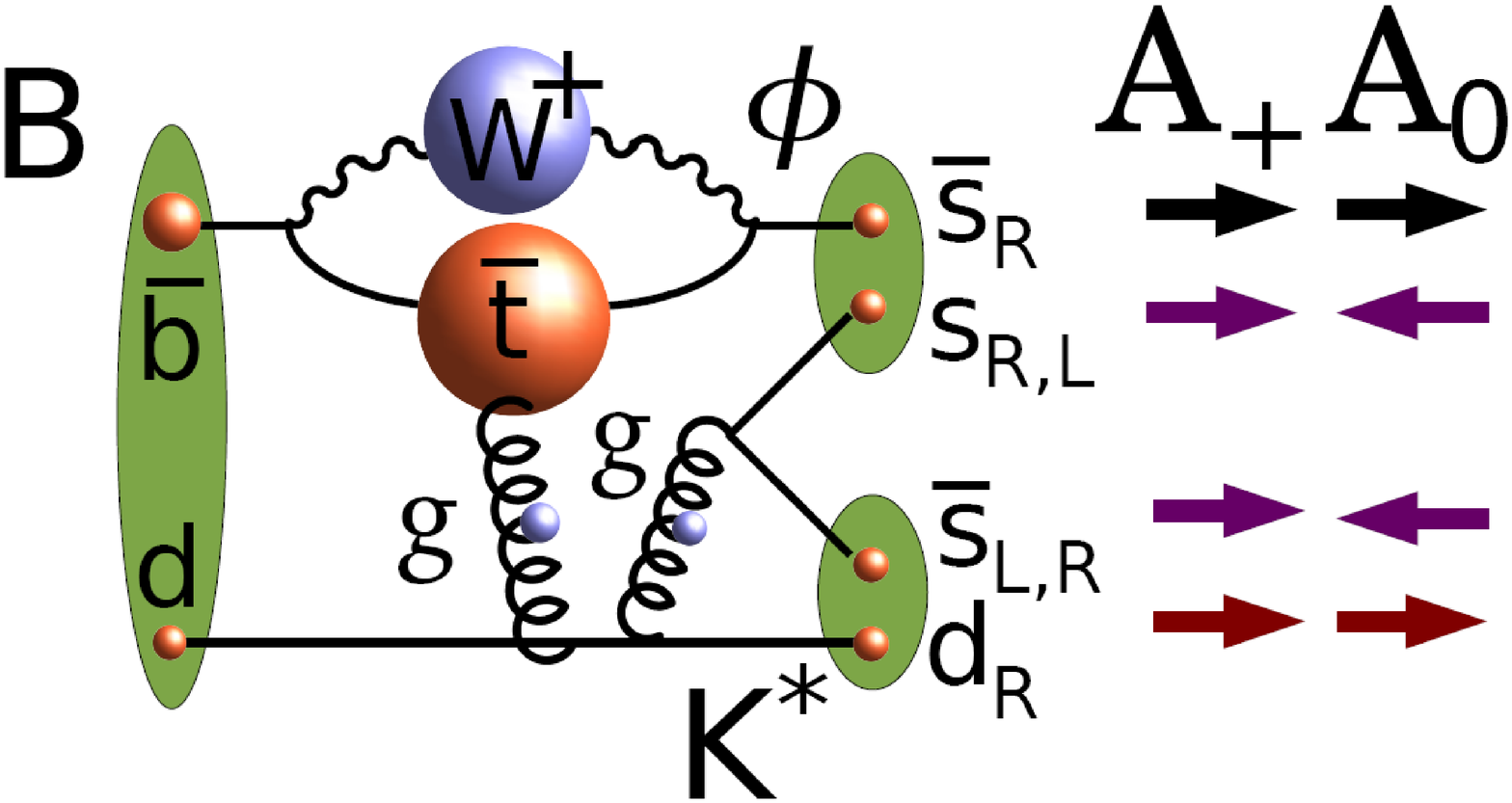}
}
\centerline{
\setlength{\epsfxsize}{0.99\linewidth}\leavevmode\epsfbox{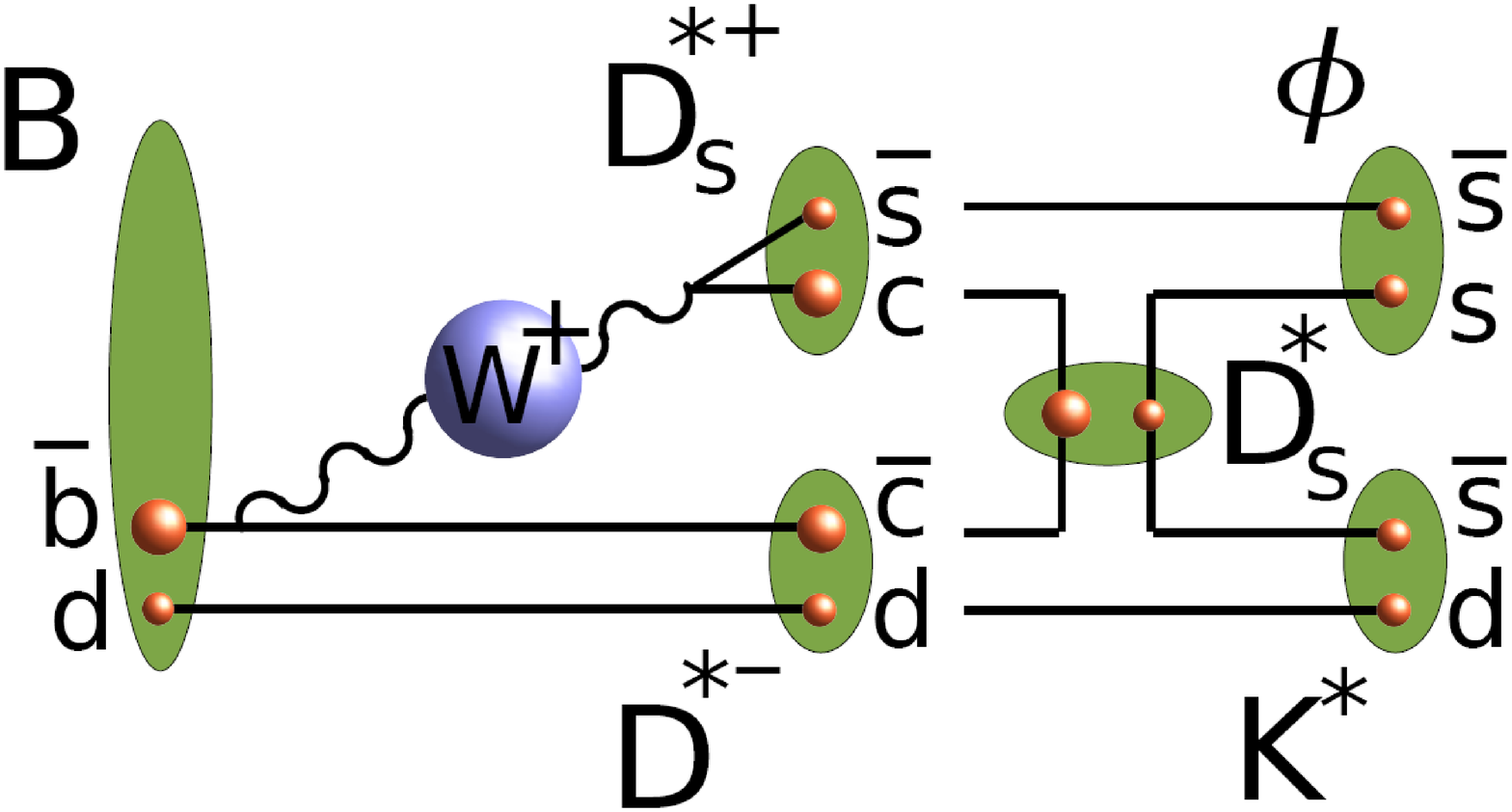}
}
\caption{Feynman diagram describing the $B\to\varphi K^{*}$  decay
with the new mechanisms within the
standard model, such as annihilation penguin~\cite{smtheory} (top)
or QCD rescattering~\cite{qcdtheory} (bottom).
\label{fig:feyn-phikst-old}
}
\end{figure}

\subsection{Discussion}

The yield of the $\varphi({K\pi})^{*\pm}_0$ contribution
is $57^{+14}_{-13}$ events
with a statistical significance of  {7.9}$\sigma$,
combining the $|A_{00}|^2$ term and the interference terms
$2{\cal R\rm e}(A_{1\lambda} A^*_{00})$, which confirms the
significant $S$-wave $K\pi$ contribution observed in
the neutral $B$-decay mode~\cite{babar:vt}.
The dependence of the interference on the $K\pi$ invariant
mass~\cite{babar:vt, Aston:1987ir, jpsikpi} allows us to reject
the other solution near ($2\pi-\phi_{\parallel},\pi-\phi_{\perp}$)
relative to that in Table~\ref{tab:results}
with significance of {6.3}$\sigma$, including systematic uncertainties.
Figs.~\ref{fig:philscann} and \ref{fig:phidscann} illustrate 
resolution of the phase ambiguity.

The $(V-A)$ structure of the weak interactions,
helicity conservation in strong interactions,
and the $s$-quark spin flip suppression in the penguin
decay diagram suggest $|A_{10}|\gg|A_{1+1}|\gg|A_{1-1}|$~\cite{bvv1}.
This expectation disagrees with our observed value of $f_L$.
We obtain the solution
${\phi_\parallel}\simeq{\phi_\perp}$ without discrete ambiguities,
which is consistent with the approximate decay amplitude hierarchy
$|A_{10}|\simeq|A_{1+1}|\gg|A_{1-1}|$.

We find that ${\phi_\perp}$ and ${\phi_\parallel}$ deviate from either
$\pi$ or zero by more than {3.1}$\sigma$ and {2.4}$\sigma$, respectively,
including systematic uncertainties. This indicates the presence of
final-state interactions not accounted for in naive factorization.
Our measurements of the six $C\!P$-violating parameters are
consistent with zero and exclude a significant part of
the physical region. We find no evidence of $C\!P$ violation
in this decay.

We have performed a full amplitude analysis and searched
for $C\!P$-violation in the angular distribution of the
$B^\pm\to\varphi K^{*\pm}$ decay.
Our results are summarized in Table~\ref{tab:results}.
Using similar techniques, we have also searched for the
$B^0\to\phi K^{*0}$ decays with the tensor
$K_3^{*}(1780)^{0}$ and $K_4^{*}(2045)^{0}$,
vector $K^{*}(1680)^{0}$, and scalar nonresonant $(K\pi)_0^{*0}$
contributions
with $K^{*0}\to K^+\pi^-$ invariant mass above 1.6 GeV.
Our results are summarized in Table~\ref{tab:results-all}.
Fig.~\ref{fig:kpimass-neutral} shows distribution of the $K\pi$ 
invariant mass in this study.
We do not find significant signal with the above resonances and
place upper limits on the $B$-decay branching fractions.

These results find substantial $A_{1+1}$ amplitude in the
$B\to\varphi K^{*}(892)$ decay and point to physics outside
the standard model
as shown in Fig.~\ref{fig:feyn-phikst-new}~\cite{nptheory},
where scalar interaction would look more consistent with 
the data, or new dynamics as shown 
in Fig.~\ref{fig:feyn-phikst-old}~\cite{smtheory,qcdtheory},
where penguin annihilation is more consistent with 
sizable $A_{1+1}$ and small $A_{1-1}$.
Any satisfactory solution to this polarization puzzle 
is expected to explain all polarization data and to be 
predictive for other experimental measurements.


\begin{acknowledgments}
I am grateful for the work of my $\babar$ colleagues who made this
contribution possible and for the excellent luminosity and machine 
conditions provided by our \pep2\ colleagues. 
I would like to thank Guglielmo De Nardo and Steve Sekula 
for discussion of the $\tau\nu$ results, Zijin Guo and Yanyan Gao
for discussion of the $\varphi K^*$ results, and 
the conference organizers for hospitality. 
This contribution was supported in part by the U.S. National Science 
Foundation and Alfred P. Sloan Foundation, and $\babar$ institutions
are supported by the national funding agencies.
\end{acknowledgments}


\begin{thebibliography}{99}


\bibitem{c}
N. Cabbibo,
Phys.\ Rev.\ Lett.\ {\bf 10}, 531 (1963).

\bibitem{km}
M. Kobayashi and T. Masakawa,
Prog.\ Theor.\ Phys.\ {\bf 49}, 652 (1973).

\bibitem{higgs}
W. S. Hou,
Phys.\ Rev.\ D {\bf 48}, 2342 (1993).

\bibitem{Isidori2006pk}
  G.~Isidori and P.~Paradisi,
  Phys.\ Lett.\  B {\bf 639}, 499 (2006).

\bibitem{Akeroyd2007eh}
  A.~G.~Akeroyd and C.~H.~Chen,
  Phys.\ Rev.\  D {\bf 75}, 075004 (2007).

\bibitem{pdg2006}
Particle Data Group,  W.-M. Yao {\it et al.}, J. Phys. G33, 1 (2006).

\bibitem{fb}
HPQCD Collaboration, A. Gray {\it et al.},
Phys.\ Rev.\ Lett.\ {\bf 95}, 212001 (2005).


\bibitem{babar-prd-btn}
$\babar$ Collaboration, B.~Aubert {\it et al.},
Phys.\ Rev.\ D {\bf 73}, 057101 (2006).

\bibitem{belle}
  K.~Ikado {\it et al.},
  Phys.\ Rev.\ Lett.\  {\bf 97}, 251802 (2006).

\bibitem{bvv1}
A.~Ali {\it et al.}, Z.\ Phys.\ C {\bf 1}, 269 (1979);
G.~Valencia, Phys.\ Rev.\ D {\bf 39}, 3339 (1989);
G. Kramer and W.F. Palmer, Phys.\ Rev.\ D {\bf 45}, 193 (1992);
H.-Y.~Cheng and K.-C.~Yang, Phys.\ Lett.\ B {\bf 511}, 40 (2001);
C.-H. Chen {\it et al.}, Phys.\ Rev.\ D {\bf 66}, 054013 (2002);
M.~Suzuki, Phys.\ Rev.\ D {\bf 66}, 054018 (2002);
A.~Datta and D.~London, Int.\ J.\ Mod.\ Phys.\ A {\bf 19}, 2505 (2004).

\bibitem{babar:vv}
$\babar$ Collaboration, B.~Aubert {\it et al.},
presented at 1st International Workshop On Frontier Science: 
Charm, Beauty And CP, Frascati, Italy (October 2002);
arXiv:hep-ex/0303020;
Phys.\ Rev.\ Lett.\ {\bf 91}, 171802 (2003);
{\bf 93}, 231804 (2004).

\bibitem{belle:phikst}
Belle Collaboration, K.-F. Chen {\it et al.},
Phys.\ Rev.\ Lett.\ {\bf 91}, 201801 (2003);
{\bf 94}, 221804 (2005).

\bibitem{belle:rhokst}
Belle Collaboration, J. Zhang {\it et al.},
Phys.\ Rev.\ Lett.\ {\bf 95}, 141801 (2005).

\bibitem{babar:rhokst}
$\babar$ Collaboration, B.~Aubert {\it et al.},
Phys.\ Rev.\ Lett.\ {\bf 97}, 201801 (2006).

\bibitem{babar:vt}
$\babar$ Collaboration, B.~Aubert {\it et al.},
Phys.\ Rev.\ Lett.\ {\bf 98}, 051801 (2007);
arXiv:0705.0398 [hep-ex]; arXiv:0705.1798 [hep-ex].

\bibitem{bvvreview2006}
A.~V.~Gritsan and J.~G.~Smith, ``Polarization in $B$ Decays''
review in~\cite{pdg2006}, J. Phys. G33, 833 (2006).

\bibitem{nptheory}
Y.~Grossman,  Int.\ J.\ Mod.\ Phys.\ A {\bf 19}, 907 (2004);
E.~Alvarez {\it et al.}, Phys.\ Rev.\ D {\bf 70}, 115014 (2004);
P.~K.~Das and K.~C.~Yang, Phys.\ Rev.\ D {\bf 71}, 094002 (2005);
C.~H.~Chen and C.~Q.~Geng, Phys.\ Rev.\ D {\bf 71}, 115004 (2005);
Y.~D.~Yang {\it et al.}, Phys.\ Rev.\ D {\bf 72}, 015009 (2005);
K.~C.~Yang, Phys.\ Rev.\ D {\bf 72}, 034009 (2005);
S.~Baek, Phys.\ Rev.\ D {\bf 72}, 094008 (2005);
C.~S.~Huang {\it et al.}, Phys.\ Rev.\ D {\bf 73}, 034026 (2006);
C.~H.~Chen and H.~Hatanaka, Phys.\ Rev.\ D {\bf 73}, 075003 (2006);
A.~Faessler {\it et al.}, Phys.\ Rev.\ D {\bf 75}, 074029 (2007).

\bibitem{smtheory}
A.~L.~Kagan, Phys.\ Lett.\ B {\bf 601}, 151 (2004);
H.~n.~Li and S.~Mishima, Phys.\ Rev.\ D {\bf 71}, 054025 (2005);
C.-H. Chen et al., Phys.\ Rev.\ D {\bf 72}, 054011 (2005);
M.~Beneke {\it et al.}, Phys.\ Rev.\ Lett.\  {\bf 96}, 141801 (2006),
arXiv:hep-ph/0612290;
C.-H. Chen and C.-Q. Geng, Phys.\ Rev.\ D {\bf 75}, 054010 (2007);
A.~Datta {\it et al.}, arXiv:0705.3915 [hep-ph].

\bibitem{qcdtheory}
C.~W.~Bauer {\it et al.}, Phys.\ Rev.\ D {\bf 70}, 054015 (2004);
P.~Colangelo {\it et al.}, Phys.\ Lett.\ B {\bf 597}, 291 (2004);
M.~Ladisa {\it et al.}, Phys.\ Rev.\ D {\bf 70}, 114025 (2004);
H.~Y.~Cheng {\it et al.}, Phys.\ Rev.\ D {\bf 71}, 014030 (2005).

\bibitem{Aston:1987ir}
LASS Collaboration,  D.~Aston {\it et al.},
Nucl.\ Phys.\ B {\bf 296}, 493 (1988);
W.~M.~Dunwoodie, private communications.

\bibitem{jpsikpi}
$\babar$ Collaboration, B.~Aubert {\it et al.},
Phys.\ Rev.\ D {\bf 71}, 032005 (2005);
Phys.\ Rev.\ D {\bf 72}, 072003 (2005).

\bibitem{babar}
\babar\ Collaboration, B.~Aubert {\it et al.},
Nucl.\ Instrum.\ Methods {\bf A479}, 1 (2002).

\bibitem{geant} S.~Agostinelli {\it et al.},
{Nucl.\ Instr.\ Meth.\xspace} A {\bf 506}, 250 (2003).

\bibitem{evtgen}
D.\ J.\ Lange,
Nucl.\ Instrum.\ Methods {\bf A462}, 152 (2001).

\bibitem{cc}
{Charge-conjugate modes are implied throughout the paper.}

\bibitem{thrust}
E.~Farhi,
Phys.\ Rev.\ Lett.\ {\bf 39}, 1587 (1977).

\bibitem{arguspdf}
  H.~Albrecht {\it et al.}  [ARGUS Collaboration],
  Phys.\ Lett.\  B {\bf 185}, 218 (1987).

\bibitem{lista}
L.\ Lista,
Nucl.\ Instrum.\ Methods {\bf A462}, 152 (2001).

\bibitem{cls}
A.\ L.\ Read,
J.\ Phys. {\bf G28}, 2693 (2002).

\bibitem{babar-note1654}
$\babar$ Collaboration, B.~Aubert {\it et al.},
arXiv:0705.1820 [hep-ex],  Submitted to Phys. Rev. D.

\bibitem{ckm}
CKMfitter collaboration, J. Charles  {\it et al.},
Eur. Phys. J. {\bf C41}, 1  (2005); http://ckmfitter.in2p3.fr/
 
\bibitem{bigPRD}
$\babar$ Collaboration, B.~Aubert {\it et al.},
Phys.\ Rev.\ D {\bf 70}, 032006 (2004).

\bibitem{f0mass}
E791 Collaboration, E. M. Aitala {\it et al.},
Phys. Rev. Lett. {\bf 86}, 765 (2001).







\end{thebibliography}
\end{document}